\documentclass[12pt]{article}
\usepackage{amsmath,epsf,amssymb,latexsym,cite}
\usepackage[ps,dvips,matrix,arrow,frame,import,curve,color]{xy}
\newif\iffigs\figstrue

\DeclareFontFamily{U}{rsf}{}
\DeclareFontShape{U}{rsf}{m}{n}{
  <5> <6> rsfs5 <7> <8> <9> rsfs7 <10-> rsfs10}{}
\DeclareMathAlphabet\Scr{U}{rsf}{m}{n}

\def\pplogo{\vbox{\kern-\headheight\kern -29pt
\halign{##&##\hfil\cr&{%\sc
\ppnumber}\cr\rule{0pt}{2.5ex}&\ppdate\cr}
}}
\makeatletter
\def\ps@firstpage{\ps@empty \def\@oddhead{\hss\pplogo}%
  \let\@evenhead\@oddhead % in case an article starts on a left-hand page
}
\def\maketitle{\par
 \begingroup
 \def\thefootnote{\fnsymbol{footnote}}
 \def\@makefnmark{\hbox{$^{\@thefnmark}$\hss}}
 \if@twocolumn
 \twocolumn[\@maketitle]
 \else \newpage
 \global\@topnum\z@ \@maketitle \fi\thispagestyle{firstpage}\@thanks
 \endgroup
 \setcounter{footnote}{0}
 \let\maketitle\relax
 \let\@maketitle\relax
 \gdef\@thanks{}\gdef\@author{}\gdef\@title{}\let\thanks\relax}
\makeatother

\def\O{\Scr{O}}
\def\C{{\mathbb C}}
\def\Q{{\mathbb Q}}
\def\P{{\mathbb P}}
\def\R{{\mathbb R}}
\def\Z{{\mathbb Z}}

\def\discrim{\operatorname{discrim}}

\def\sign{\operatorname{sign}}

\def\Gl{\operatorname{GL}}

\def\p{\partial}

\def\cL{{\Scr L}}

\def\ff#1#2{{\textstyle\frac{#1}{#2}}}

\def\ev{\mathbf{e}}
\def\mv{\mathbf{m}}
\def\nv{\mathbf{n}}
\def\qv{\mathbf{q}}
\def\Qv{\mathbf{Q}}
\def\Qc{\mathcal{Q}}
\def\rv{\mathbf{r}}
\def\K{\mathcal{K}}
\def\EC{\mathcal{\chi}}
\def\U{\mathcal{U}}
\def\ECn{\EC_{\nv}}
\def\Kd{{{\K}^\vee}}
\def\N{\mathcal{N}}
\def\M{\mathcal{M}}
\def\Mn{\M_{\nv}}
\def\la{\langle}
\def\ra{\rangle}
\def\As{{a_1\cdots a_s}}

\def\tp{{\theta^+}}
\def\tm{{\theta^-}}
\def\tpm{{\theta^\pm}}
\def\tbp{\overline{\theta}^+}
\def\tbm{\overline{\theta}^-}
\def\tbpm{\overline{\theta}^\pm}

\def\Phib{\overline{\Phi}}
\def\sigmab{\overline{\sigma}}
\def\Sigmab{\overline{\Sigma}}
\def\lambdab{\overline{\lambda}}
\def\Lambdab{\overline{\Lambda}}
\def\Wt{{\widetilde{W}}}
\def\Wte{{\widetilde{W}_{\text{eff}}}}
\def\Orb{\C^3/\Z_{N(a,b,c)}}

\def\Orbt{\C^3/\Z_{(2N+1)(2,2,1)}}

\def\PPROD{{\prod_{Q_i^a > 0} \xi_i^{Q_i^a}}}
\def\MPROD{{\prod_{Q_i^a < 0} \xi_i^{-Q_i^a}}}
\def\EUL{{\prod_{d_i<0} \xi_i^{-d_i-1} }}
\def\MEUL{{\prod_{d_i-Q_i^a<0} \xi_i^{Q_i^a-d_i-1} }}

\begin{document}
\setcounter{page}0
\def\ppnumber{\vbox{\baselineskip14pt
\hbox{DUKE-CGTP-05-01}
\hbox{hep-th/0501238}}}
\def\ppdate{\today} \date{}

\title{\LARGE The Coulomb Branch in Gauged Linear Sigma Models\\[10mm]}
\author{
Ilarion V. Melnikov and M. Ronen Plesser\\[3mm]
\normalsize Center for Geometry and Theoretical Physics \\
\normalsize Box 90318 \\
\normalsize Duke University \\
\normalsize Durham, NC 27708-0318
}

{\hfuzz=10cm\maketitle}

\def\Large{\large}
\def\LARGE{\large\bf}

\vskip 1cm

\begin{abstract}
We investigate toric GLSMs as models for tachyon condensation in type
II strings on space-time non-supersymmetric orbifold singularities.
The A-model correlators in these theories satisfy a set of relations
related to the topology of the resolved orbifold.  Using these
relations we compute the correlators and find a non-trivial chiral
ring in the IR, which we interpret as supported on isolated Coulomb
vacua of the theory.
\end{abstract}

\vfil\break

%%%%%%%%%%%%%%%%%%%%%%%%%%%%%%%%%%%%%%%%%%%%%%%%%%%%%%%%%%%%%%%%

\section{Introduction}  

The study of localized closed string tachyons, first undertaken by
Adams, Polchinski and Silverstein in \cite{APS:tach}, has by now a
venerable history.  The basic picture found by these authors and
others \cite{Vafa:tachyons,HKMM:tachyons,MNP:loctac, Sarkar:tach, MN:flip} is
that the orbifold singularity is (at least partially) resolved by
tachyon condensation.  A nice recent review of this and related
matters is given in \cite{HMT:tachrev}. However, as pointed out by
Martinec and Moore \cite{MM:decay}, and explored further by Moore and
Parnachev \cite{MP:McKay}, this cannot be the entire story.  As these
authors show, if one considers the $D$-brane charges in the orbifold
theory as labeled by equivariant $K$-theory of the orbifold, then,
naively, one might conclude that these charges {\em disappear} as
tachyon condensation drives the system to flat space.  These missing
charges can be associated to branes on the isolated Coulomb vacua that
are found in the Gauged Linear Sigma Model (GLSM) description of this
tachyon condensation.  

We follow up on this finding by studying the closed string sector of the
GLSM, and we find that the missing $D$-brane charges and their
restoration by the Coulomb branch have analogues in the closed string
topological observables.  We compute the correlators of the
topological $A$-twisted model of the GLSM, and find that the isolated
Coulomb vacua responsible for carrying the missing $K$-theory charges
also support a non-trivial chiral ring.  To make our explicit
calculations possible, we restrict attention to a class of tractable
models, as described below.   We will find a rather
rich set of examples amenable to our approach.

We consider localized tachyons of type II strings on space-time
non-supersymmetric orbifolds of the form $\Orb$. The remarkable
tractability of these backgrounds, which closely parallels the case of
open string tachyons associated to unstable configurations of
$D$-branes\cite{Sen:bigrev}, makes this an ideal laboratory for the
study of closed string tachyon condensation.  These theories can be
explicitly constructed as orbifolds of a free CFT\cite{DFMS:orb} which
leads to the identification of the localized tachyon vertex operators
as certain twisted sector operators.  While the GSO projection removes
the bulk closed string tachyon, it leaves some of these twisted-sector
operators in the theory.  In addition, by suitably tuning $\alpha'$
and $g_s$, string corrections can be made small for a tachyon
condensate sufficiently large to display a marked departure from the
initial, unstable string vacuum.  Thus, we can study tachyon
condensation at string tree level, and we may hope that this captures
some essential features of the endpoint of tachyon condensation.

Time evolution is difficult to study even for these simple backgrounds,
and calculating the tachyon potential can be quite a
challenge\cite{DIR}.  Instead, we study the easier problem of RG
flow in the space of $d=2$ theories: since there is a natural relevant
perturbation of the orbifold CFT associated to the localized tachyon,
and the IR fixed point of the flow resulting from this perturbation
will be a solution of the (tree-level) string equations of motion, one
might hope that the IR fixed point correctly describes the endpoint of
tachyon condensation.  Although it has been argued that RG flow
correctly interpolates between different string
backgrounds\cite{HMT:tachrev}, and this has been shown to hold in
several examples, this statement may be too strong in general. Still,
it is certainly true that RG flow provides an accessible and
interesting interpolation between string vacua.  We will adopt this
perspective and study the IR fixed points of RG flows where the UV
fixed point is the orbifold CFT and the relevant perturbation
corresponds to condensing the tachyon.

The orbifold CFT is actually a superconformal theory, invariant under an
$\N=(2,2)$ SUSY algebra.  The relevant deformation corresponding
to a generic tachyon completely breaks this supersymmetry, and thus,
the powerful constraints of $\N=(2,2)$ SUSY cannot be used to
constrain the RG flow.  However, the deformations corresponding to
tachyons that are chiral primary operators do preserve $\N=(2,2)$
supersymmetry.  Furthermore, the most relevant tachyonic operator,
i.e. the one with the largest negative mass-squared, and thus
corresponding to the fastest growing instability, is a chiral primary
operator\cite{MNP:loctac,LS}.  The RG flows associated to these chiral
primary tachyons are much more tractable, and in this paper we will
work with these flows.  We may think of this as either fine-tuning
away the other tachyons, or just simplifying the analysis by studying
the most unstable mode, which, after all, should grow exponentially
faster than the others, and thus be the dominant effect for a
substantial (RG) time.  

Although we know the orbifold SCFT and the relevant perturbation of
interest, the RG flow is still too difficult to study directly, mainly
because it involves perturbing the theory by a twisted-sector
operator.  However, since these orbifolds are toric, each can be
realized as the low energy limit of a simpler massive theory: a GLSM
with zero superpotential.  By a suitable choice of scales and
parameters, we can ensure that, barring the emergence of a non-trivial
separatrix, the RG flow of the GLSM passes arbitrarily near to the
orbifold SCFT fixed point for many decades of the RG scale, and it
then follows the relevant perturbation associated to tachyon
condensation.  

Our tool for exploring these models will be a study of the correlators
in their topological twisted version.  These encode properties of the
GLSM that are interesting and yet computable.  Following the seminal
work of Witten \cite{W:phases}, Morrison and Plesser provided a set of
toric methods that allow us to compute certain RG invariants ({\em
i.e.} topological observables) in these GLSMs \cite{MP:summing}.  As
we will show, several surprises emerge upon taking a closer look at
GLSMs that correspond to space-time non-supersymmetric backgrounds.
The most impressive of these is that the instanton computation of
\cite{MP:summing} describes these correlators only in some of the
phases.  In others, the correlators cannot be computed by expanding in
the instantons about a Higgs vacuum.  These phases are additionally
characterized by the existence of isolated Coulomb vacua far from the
Higgs branch, and we conjecture that these support the missing
correlators. Although we have not been able to confirm this by an
explicit computation, we are able to show that the form of the
correlators strongly suggests a Coulomb branch interpretation.

What this finding tells us about the physics of the decay process is
not clear.  It is difficult to understand what a set of isolated
vacua of the world-sheet theory could mean in space-time.  It may be
relevant in this context that the models in question all involve
non-compact ``compactification'' spaces.  The picture of tachyon decay
advocated in \cite{APS:tach} describes an expanding bubble, in the
interior of which the local structure is that of the resolved space,
joined to the outside by some complicated structure on the bubble
wall.  It seems natural to guess that the Coulomb branch represents,
at least at the level of the topological model, the structures that
``disappear'' to infinity in the RG flow.  In the expanding bubble
picture, these structures would live on the bubble wall (or outside
it). 

The rest of the paper is organized as follows:  in section \ref{s:GLSM}
we briefly recall some salient features of the GLSM relevant for tachyon
condensation; in section \ref{s:Example} we study a simple two-parameter
example of $\Orbt$, which encapsulates much of the essential physics of 
our results; in section \ref{s:MR} we discuss the quantum cohomology 
relations and the constraints they place on the topological correlators; 
in section \ref{s:General}, we use these constraints to generalize the 
analysis of the example,  and we wrap up with a discussion in section 
\ref{s:Discussion}.
%%%%%%%%%%%%%%%%%%%%%%%%%%%%%%%%%%%%%%%%%%%%%%%%%%%%%%%%%%%%%%%%%
\section{The structure of the Gauged Linear Sigma Model} \label{s:GLSM}
The power of the GLSM approach to $\Orb$ theories takes root in the 
$\N= (2,2)$ world-sheet supersymmetry and its close relation to toric
geometry.  In this section we will outline some of the models' basic 
features from those perspectives.
%%%%%%%%%%%%%%%%%%%%%%%%%%%%%%%%%%%%%%%%%%%%%%%%%%%%%
\subsection{$\N=(2,2)$ SUSY and the GLSM Lagrangian} 
We will present the field content and Lagrangian of the GLSM in  $d=2$,
$\N = (2,2)$ superspace.  Labeling the four anticommuting coordinates 
by $\tpm$,$\tbpm$, the GLSM involves the following fields.
The $n$ chiral matter fields $\Phi^i$, $i=1,\ldots,n$ have a superspace 
Taylor expansion 
\begin{equation}
 \Phi^i = \phi^i + \sqrt{2}\left( \tm\psi_-^i +\tp\psi_+^i\right) -2\tm\tp F^i + \ldots,
\end{equation}
with $\phi^i$ a complex scalar, $\psi_\pm^i$ right/left-moving 
Weyl fermions, and $F^i$  an auxiliary complex field needed for 
off-shell closure of the SUSY algebra. The higher components are 
determined by the constraints in terms of the presented components.
These matter fields are minimally coupled to $u$ abelian 
gauge fields with charges $Q^a_i$, $a=1,\ldots,u$.  The gauge 
fields $v_{a,\mu}$ reside in vector multiplets $V_a$ which 
(in Wess-Zumino gauge) have the form
\begin{eqnarray}
V_a & = & \tp\tbp v_{a,+} - \tm\tbm v_{a,-} + 
\left[ \sqrt{2} \tm\tbp\sigma_a \right.\nonumber \\
~ & ~ & \left. + 2i \tm\tp\left(\tbp\lambdab_{a,+} + \tbm\lambdab_{a,-}\right) 
               -\tm\tp\tbm\tbp D_a  + \text{c.c.} \right],
\end{eqnarray}
with $\sigma_a$ a complex scalar, $\lambda_{a,\pm}$ left/right 
Weyl fermions, and $D_a$ an auxiliary real field.   The gauge field 
strength resides in the third and last type of multiplet that we will need:  
the {\em twisted} chiral multiplet $\Sigma$ with the superspace expansion
\begin{equation}
\Sigma = \sigma + i\sqrt{2} \tp\lambdab_+ - i\sqrt{2}\tbm\lambda_- 
         + \sqrt{2} \tp\tbm\left(D -i f_{01}\right) + \ldots,
\end{equation}
where $f_{a,01} = \p_0 v_{a,1} - \p_1 v_{a,0}$ is the field strength, and
$\sigma_a$, $\lambda_{a,\pm}$, and $D_a$ are as above.  (Super)gauge
transformations leave $\Sigma_a$ invariant, while $\Phi^i$ and $V_a$
transform according to 
\begin{eqnarray}
\Phi^i & \to & \exp\left(\sum_a Q^a_i \Lambda_a\right) \Phi^i \nonumber,\\
  V_a & \to & V_a - (\Lambda_a + \Lambdab_a)/2,
\end{eqnarray}
where $\Lambda_a$ is a chiral superfield.

We define the GLSM at a scale $\mu$ by a Lagrange density $\cL^\mu$ given
by a sum of three terms, the K\"ahler term $\cL_K^\mu$, the superpotential 
$\cL_W^\mu$, and the {\em twisted superpotential} $\cL_\Wt^\mu$.  
We take the K\"ahler term to be 
\begin{equation}
\cL_K^\mu = \int d^4\theta\left( - \frac{1}{4} \sum_{i=1}^n \Phib^i 
                        \exp\left(2 \sum_{a=1}^{u} Q^a_i V_a \right) \Phi^i
                      + \frac{1}{4 \mu^2 g(\mu)^2} \sum_{a=1}^{u} \Sigmab_a \Sigma_a\right),
\end{equation}
where $g(\mu)$ is the dimensionless coupling of the gauge theory.  The models of 
interest to us will have the superpotential term $\cL_W$ set to zero, and for 
reasons that will be made clear below, we will call such theories {\em toric} 
GLSMs.  Finally, the tree-level twisted superpotential is given by
\begin{equation}
\cL_\Wt^\mu = \left[-\frac{i}{2\sqrt{2}}  \int d\tp d\tbm \sum_{a=1}^{u} \Sigma_a \tau^a(\mu)\right]  + \text{c.c.}.
\end{equation}
The $\tau^a(\mu) = i r^a(\mu) + \ff{\theta^a}{2\pi}$ are the parameters of the 
model.  Each $\tau^a$ is a  combination of the Fayet-Iliopoulos (F-I) term $r^a$ and 
the $\theta$-angle $\theta^a$. It is useful to define single-valued 
parameters $q_a = e^{2\pi i \tau^a}.$ 

$\cL^\mu$ provides a good description of the degrees of freedom and their
interactions at external momenta of order $\mu$. The GLSM is asymptotically
free, and the low energy theory is strongly coupled in terms of these degrees of
freedom.  For more details about the structure of $\N=(2,2)$ SUSY and 
the GLSM Lagrangian the reader should consult \cite{W:phases,WB:susy}. 
%%%%%%%%%%%%%%%%%%%%%%%%%%%%%%%%%%%%%%%%%%%%%%%%%%%%%
\subsection{The phases of the GLSM, I} \label{ss:GLSMP}
Although quantum effects substantially modify the low energy 
description of the theory, it pays to consider the classical moduli space of
the GLSM.  The classical moduli space is the set of supersymmetric vacua, 
para\-metrized, as usual, by the zeroes of the classical scalar potential 
modulo the gauge group.  The toric GLSM has the scalar potential 
\begin{equation}
\label{eq:scapot}
U(\phi^i,\sigma_a) = 2 \sum_{i,a,b} |\phi^i|^2 Q_i^a Q_i^b \sigma_a \sigmab_b 
                     + \frac{1}{2 \mu^2 g(\mu)^2} \sum_a \left(D^a\right)^2,
\end{equation}
with
\begin{equation}
D^a = \mu^2 g(\mu)^2 \left( \sum_i Q_i^a|\phi^i|^2 - r^a \right).
\end{equation}
The classical analysis is straightforward.  Despite the lack of a dynamical
two-dimensional photon or spontaneous symmetry breaking in $d=2$,  we 
can describe the classical physics as the Higgs
mechanism.\footnote{Going beyond the classical theory, one could imagine this 
as the starting point for a semi-classical Born-Oppenheimer analysis of the 
vacuum structure.} Generically, the gauge group is completely Higgsed,
$\sigma_a$ are all massive, and solving for $D^a = 0$ for
all $a$ modulo the gauge group, one finds that the moduli space is a 
complex dimension $d=n-u$ toric variety, whose geometric properties 
are independent of the $\theta$-angles and
vary smoothly with the $u$ F-I parameters $r^a$.  
The geometry degenerates as certain cycles shrink to zero size along 
co-dimension one walls in the $\R^{u}$ space spanned by the $r^a$.
This process is a physical realization of the familiar blow-ups and 
blow-downs of algebraic (for our purposes toric) geometry. The regions 
of the parameter space separated by the walls are termed the 
{\em phases} of the GLSM.  The phases turn out to be the
full-dimensional cones in the secondary fan associated to the
toric fan of the GLSM.

When the F-I terms $r^a(\mu)$ 
are deep in the interior of a particular
phase $\K$, corresponding to a classical moduli space $V$, then the light 
fields correspond to those of the NLSM with a target space that is 
topologically equivalent to $V$.  Furthermore, these light fields are 
weakly coupled to the massive fields.  Thus, by choosing $r^a(\mu)$ 
sufficiently deep in $\K$, we can ensure that the RG trajectory of the
GLSM passes arbitrarily close to that of a NLSM with target space $V$.
In particular, by appropriately choosing a GLSM and $r^a(\mu)$ we can
ensure that the high energy theory is well approximated by an orbifold
theory of the form $\Orb$.  As explained in \cite{MM:decay}, the necessary
limit is to choose a fiducial scale $\mu^*$ and send $g(\mu^*) \to \infty$,
while holding $r^a(\mu^*)$ fixed (and in the correct phase).  

On the other hand, when $r^a(\mu)$ approaches a phase boundary, the
description as a NLSM with target space $V$ seems to break down.  As 
we discuss in the next section, this is not entirely as it seems.

%%%%%%%%%%%%%%%%%%%%%%%%%%%%%%%%%%%%%%%%%%%%%%%%%%%%%
\subsection{Quantum moduli space and the one loop $\beta$-function}
Quantum effects modify the classical phases picture in several 
important ways.  The most basic is the one loop renormalization 
of the F-I parameters:
\begin{equation}
\mu \frac{\p}{\p \mu} r^a = \frac{1}{2\pi}\sum_i Q^a_i.
\end{equation}
World-sheet supersymmetry ensures that this result is uncorrected in 
perturbation theory.  More precisely, holomorphy ensures that $\tau^a$ 
is uncorrected beyond the shift above.

It is easy to show that by a $\Gl(u,\Z)$ transformation one 
can always choose a basis for the $Q^a_i$ such that 
$\sum_i Q^1_i = \Delta$, and $\sum_i Q^a_i = 0$ for 
$a=2,\ldots,u$.  Such a change of basis leaves the classical
moduli space unchanged,  and it preserves the IR 
physics, since it only changes the structure of the gauge field
kinetic terms---an irrelevant 
K\"ahler deformation.  Let us assume that such a basis is chosen.  
If $\Delta = 0$, then $r^1$ does not run, and each phase is
bi-rationally equivalent to a toric Calabi-Yau manifold.  
The resulting string theory ``compactification'' is supersymmetric 
and fairly well understood \cite{MP:summing, HV:mirsym}.  We will be
interested in models where $\Delta \neq 0$.  In that case,
at low energies the model is driven to $r^1 = -\infty$ if
$\Delta > 0$ and to $r^1 = +\infty$ if $\Delta <0$.\footnote{Obviously,
$r^1$ runs and should not really be considered a parameter of the theory.
Similarly, the corresponding $\theta$-angle $\theta^1$ is not a parameter
either:  the RG running is accompanied by an anomaly in the axial $R$-symmetry,
which breaks the $U(1)_A$ to $\Z_{|\Delta|}$.  This anomaly allows us to
to set $\theta^1$ to a $\Delta$-th root of unity.}  As discussed
in \cite{Vafa:tachyons, MM:decay, MNP:loctac}, this is 
the RG flow which corresponds to tachyon condensation
in space-time.  It is convenient to express this running in
terms of $q_1$:
\begin{equation}
q_1(\mu) = q_1(\mu_0) \left(\frac{\mu_0}{\mu}\right)^\Delta.
\end{equation}
While $q_1$ runs, the $q_a$ for $a > 1$ are RG invariants and
parametrize exactly marginal deformations of the GLSM. 

In addition, quantum effects change the nature of the
classical singularities \cite{W:phases,MP:summing}.  Recall that these singularities
in the low energy NLSM description were due to the appearance
of massless $\sigma^a$ along co-dimension $1$
walls in the space of $r^a$.  Each wall is associated to
the un-Higgsing of a particular gauge group.  Suppose the charges
of the un-Higgsed gauge group are $Q_i$, the complex scalar is $\sigma$,
and the complex parameter is $q$.  Furthermore, suppose that $q$ is tuned to
the vicinity of the classical singularity, $|q| = 1$.  When 
$\sum_i Q_i \neq 0$, quantum effects generate a potential for 
$\sigma$, and thus smooth out the classical singularity.  On the 
other hand, if $\sum_i Q_i =0$, then for a particular value of $q$,
there is no potential generated for $\sigma$, and thus a continuous
Coulomb branch emerges.  This Coulomb branch is parametrized by
the expectation value of $\sigma$.  The semi-classical analysis
leading to this result is valid when $|\sigma|$ is large and the
other gauge groups are Higgsed with large F-I terms.  This 
analysis shows that the {\em singular locus} of the theory, i.e. the
values of the $q^a$ where the low energy NLSM description breaks down,
is a {\em complex} co-dimension one subvariety in the space of the 
GLSM parameters $q^a \in \C^u$.  Thus, all the phases are connected 
by paths along which the low energy NLSM description is non-singular. 
%%%%%%%%%%%%%%%%%%%%%%%%%%%%%%%%%%%%%%%%%%%%%%%%%%%%%
\subsection{The Quantum Coulomb branch} \label{ss:coulombvac}
In addition to the continuous Coulomb branch which exists on the
singular locus, when $\Delta\neq 0$, the GLSM possesses another set of Coulomb
vacua.  These are isolated vacua, whose locations vary continuously
with the GLSM parameters.  To study the Coulomb branches in more
detail, it pays to return to the classical scalar potential of eqn. (\ref{eq:scapot}).
The form of the potential indicates that when the $\sigma_a$ acquire
non-zero expectation values, they give masses to some or all of the
$\phi^i$.  We may integrate out these massive chiral matter fields
to obtain an effective action for the $\sigma_a$  appropriate at 
sufficiently low energies. In the case that all the matter fields
are massive, $\N=(2,2)$ supersymmetry combined with  't Hooft anomaly 
matching determine the effective twisted superpotential at scale $\mu$ 
to be \cite{MP:summing}
\begin{equation}
\Wte = - \frac{1}{4 \pi \sqrt{2}} \sum_{a=1}^{u} \Sigma_a 
        \log\left[ \prod_{i=1}^n \left(\frac{1}{\exp(1) \mu} 
                   \sum_{b=1}^{u} Q_i^b\Sigma_b\right)^{Q_i^a} /q_a\right].
\end{equation}
When is this effective description valid? It is not valid at
high energies when $r^a(\mu)$ are deep in the interior of a phase.  Here 
we know that there are light $\phi$ degrees of freedom that are certainly
missed in the $\Wte$ description.  However, it is perfectly plausible for 
a Coulomb branch to emerge at low energies or at the singularities on the 
phase boundaries.  Indeed, the former is necessary to match the UV and IR
computations of the Witten index, and the latter provides an explanation
of how the GLSM resolves the NLSM singularities.

The equations of motion that follow from $\ff{\p\Wte}{\p \sigma^a} = 0$
are
\begin{equation}
\label{eq:teom}
\prod_i \left(\frac{1}{\mu} \sum_b Q_i^b \sigma_b\right)^{Q_i^a} = q_a.
\end{equation}
We may write these as polynomial relations among the $\sigma_a$: 
\begin{equation}
\label{eq:teompoly}
\prod_{i|Q_i^a > 0} \left(\sum_b Q_i^b \sigma_b\right)^{Q^a_i} - 
\mu^{\Delta_a} q_a \prod_{i|Q_i^a < 0} \left(\sum_b Q_i^b \sigma_b\right)^{-Q^a_i}=0,
\end{equation}
where, as before, $\Delta_a = \sum_i Q_i^a$.  To be precise, the last two
equations are equivalent only if the two products in the second equation have no
common factors.  This will be an assumption in what follows,  but we 
will see that it is satisfied in a wide class of examples. 

For some purposes it is convenient to choose a standard basis of $Q_i^a$ 
such that $\sum_i Q_i^1 = \Delta$ and $\Delta_a = 0$ for $a > 1$.
Making this choice, and letting
\begin{eqnarray}
\label{eq:teomdefs}
\omega_a & = & \sigma_a / \sigma_1 ~~~, a>1,\nonumber\\
\zeta_i   & = & Q^1_i + \sum_{a>1} Q_i^a\omega_a, \nonumber\\
s(\omega_2,\ldots,\omega_u) &=& \prod_i \zeta_i^{-Q_i^1},
\end{eqnarray} 
the equations can be written as 
\begin{eqnarray}
\label{eq:teomsols}
P_a(\omega_2,\ldots,\omega_u) & = & 
\prod_{Q_i^a > 0} \zeta_i^{Q_i^a} - q_a \prod_{Q_i^a <0} \zeta_i^{-Q_i^a} = 0 ~~~a>1,
\nonumber\\
\sigma_1^\Delta & = & \mu^\Delta q_1 s(\omega_2,\ldots,\omega_u), \nonumber\\
\sigma_a & = & \omega_a \sigma_1, ~~~a>1.
\end{eqnarray} 
The case $u=2$ is simple, and since we will study it in detail below, 
we will specialize the above expressions to it.  There is a single 
$\omega$, which takes on the $\deg(P)$ values of the roots of a single 
polynomial $P(\omega)$.  The degree of $P(\omega)$ is given by 
$\deg (P) = \sum_{Q_i^2>0} Q_i^2$.  For $u=2$ and generic
$q_2$ there are $|\Delta| \deg(P)$ distinct $\sigma$-vacua.  It is
no accident that the $\sigma$-vacua come in families of $|\Delta|$ vacua
related by roots of unity:  the axial $U(1)$ $R$-symmetry, which is
anomalous in the high energy description of the GLSM, is 
spontaneously broken in these Coulomb vacua to $\Z_{|\Delta|}$.
Note that the condition that eqn. (\ref{eq:teom}) is equivalent to 
eqn. (\ref{eq:teompoly}) is that $P(\omega)$ is irreducible over $\Z$.

We have seen how the twisted superpotential produces the isolated Coulomb
vacua.  Of course, it may also be used to study the emergence of the continuous
Coulomb branch.   The $\Wte$ given above was derived under the assumption that
all of the matter fields are massive.  This need not be the only possibility.
It is possible to find vacua where  some of the $\phi^i$ are massive, while 
others have expectation values that, in turn,  give masses to  some of 
the $\sigma_a$.   In particular, working in the basis of $Q^a_i$ given
above, one can see that there may be vacua where the gauge group corresponding to $Q^1_i$ is 
Higgsed, while the rest of the $\sigma_a$ are massless. The
resulting effective potential for the $\sigma_a$, $a>1$ will be a set
of $u-1$ homogeneous equations with $u-1$ parameters,  leading to
an implicit form for a component of the singular locus.  When $q_a$
are tuned to this variety, a continuous Coulomb branch will arise, with
$\sigma_1 = 0$, and $\sigma_a$ large for $a >1$.  

It is important not to confuse the different types of Coulomb vacua, so
we will end by re-iterating some of the differences between the two.
The isolated Coulomb vacua arise only when $\Delta \neq 0$.  They are
found in the IR phases of the GLSM, where they provide a set of
supersymmetric vacua  that vary smoothly with the parameters $q_a$.
On the other hand, continuous Coulomb vacua only arise in GLSMs where
it is possible to un-Higgs a gauge group with charges satisfying $\sum_i Q_i =0$.
They occur only on the singular locus of the model.

In this paper, we will be largely concerned with the isolated Coulomb
vacua, so unless indicated otherwise, ``Coulomb branch'' will mean
the set of these vacua.

\subsection{The phases of the GLSM, II}
In summary, we find the following picture for the GLSM phases.  By
suitably tuning $g(\mu^*)$ and $r^a(\mu^*)$ at some fiducial
scale $\mu^*$, we ensure that the GLSM is in a phase where its low 
energy behavior is well approximated by a NLSM with a non-compact target 
space $V$ for many decades of the RG scale $\mu$.\footnote{The case of 
most interest to us will be $V \simeq \Orb$.} The GLSM is well described by a
purely ``Higgs branch'' description, again in the sense that for many
decades of $\mu$ the physics is insensitive to the eventual emergence 
of a Coulomb branch and a new Higgs branch.

At low energies, the high energy degrees of freedom
of the GLSM become strongly coupled, and a different effective description
is needed.  Despite the asymptotic freedom of the GLSM, the IR fixed point 
may, to some extent, be described by the GLSM degrees of freedom.  On the
one hand, the IR Higgs branch is again weakly coupled, since after
passing through a strongly coupled region ($r^a(\mu) \approx 0$), the
$r^a(\mu)$ are again deep in the interior of another phase, so the
statements about light fields corresponding to a NLSM with target space
$V'$ still hold.  On the other hand, the Coulomb branch can be described
by the effective twisted superpotential, whose form is completely fixed
by 't Hooft anomaly matching and supersymmetry.

A final point we would like to highlight is the question of the
separation between the isolated Coulomb and Higgs vacua. 
Essentially, we are in the usual trouble:  a SUSY theory is quite
revealing of its zero energy states, but it keeps non-zero energy
information mostly to itself, and questions about energy barriers 
between various vacua remain difficult.  Naively, one may argue that the $D$-terms of the
Higgs branch are large if evaluated on the isolated Coulomb branch.
Or, one may feel that a better argument is that the two sets of vacua 
are well separated in field space.\footnote{Note
that while using classical $D$-terms prompt one to argue that the isolated
Coulomb vacua are separated from the Higgs vacua, it would also suggest
that the continuous Coulomb branch may be connected to the Higgs branch.}
However, both arguments are affected by the renormalization of the K\"ahler
terms in the theory.  Does this renormalization qualitatively change the 
physics?  While it is tempting to say that only the development of a 
singularity could cause the branches to come arbitrarily close in field space 
and/or decrease the energy barriers between them to be arbitrarily small, 
we cannot make a rigorous statement about this.

%%%%%%%%%%%%%%%%%%%%%%%%%%%%%%%%%%%%%%%%%%%%%%%%%%%%%%%%%%%%%%%%%%
\subsection{The topological $A$-twist and instanton sums} \label{s:toptwist}
Our understanding of quantum effects is not so complete that
we can directly compute any quantities along the RG flow.  
Our strategy is to follow an old idea of Witten:  working on a world-sheet
with Euclidean signature, we can use a 
related topological (and hence RG-invariant) theory to compute
a set of RG-invariants in the original theory.  These RG invariants
can then be used to probe non-perturbative aspects of the IR physics.
The background for this approach is well covered in 
\cite{W:phases,W:mirtop,MP:summing}, and we will content ourselves with 
a cursory treatment of it here.

The existence of this topological theory, the so-called $A$-model, 
follows from the structure of the $U(1)_R \times U(1)_L$ $R$-symmetry
subgroup of the $\N=(2,2)$ supersymmetry.  The $R$-symmetry 
leaves the $\Phi^i$ and $V_a$ superfields invariant, but 
rotates the $\tpm$ with charges $Q_R(\tp) = Q_L(\tm) = 1$,
$Q_R(\tm) = Q_L(\tp) = 0$.  All other charges are determined by this
choice, and, in particular, we see that 
\begin{equation}
Q_R(\sigma_a) = -Q_L(\sigma_a) = 1.
\end{equation}
The axial $U(1)_A$ subgroup with charges $Q_A = \ff{1}{2}\left(Q_R - Q_L\right)$
is anomalous, with 
\begin{equation}
\Delta Q_A = \sum_{i,a} n_a Q_i^a,
\end{equation}
where
\begin{equation}
n_a = - \frac{1}{2\pi} \int_\Sigma f_a
\end{equation}
is the instanton number associated to the $a$-th gauge group on
the world-sheet $\Sigma \simeq \P^1$.\footnote{Note that $Q_i^a$ and $r^a$ are naturally
thought of as vectors in $\R^u$, and $n_a$ is a vector in the dual
space $\left(\R^u\right)^\vee$.}  
The vector $U(1)_V$ is non-anomalous and can
be used to twist the theory along the lines of \cite{W:phases}.
We modify the spin connection $J_T$ by defining a new Lorentz 
current $J_T' = J_T + J_V$.  This is the $A$-twist.  The
$B$-twist, associated to twisting by the anomalous axial symmetry does
not give rise to a well-defined theory when $\Delta\neq 0$.

The twisted theory has several important properties, which we now review.
First, twisting renders a linear combination of the $\N=(2,2)$ 
an anticommuting (world-sheet) scalar operator $\Qc$, which can be
used as a BRST-like operator to project the observables of the theory
onto its cohomology---the topological observables.  These topological 
observables decouple from $\Qc$-exact operators in correlation functions
and are nothing other than the twisted versions of chiral 
operators of the untwisted theory.  In the GLSM, the set of local observables
in the $\Qc$-cohomology is spanned by powers of $\sigma_a$.
The $U(1)_A$ remains an anomalous symmetry
of the theory, with $\Delta Q_A$ modified to be 
\begin{equation}
\label{eq:ghostnum}
\Delta Q_A = d + \sum_{i,a} n_a Q_i^a,
\end{equation}
where, as before, $d= n-u$.
$Q_A$ is called {\em the ghost number} in analogy with the usual BRST ghost 
number.

Second, one can show that a perturbation of the world-sheet metric corresponds
to modifying the action by a $\Qc$-exact term, so that correlation functions of
topological observables are insensitive to the world-sheet metric, and hence,
are truly topological.  It immediately follows that the correlation functions
are independent of the positions of the operator insertions on the world-sheet.

In addition, a perturbation of the coupling $g(\mu)$ corresponds to a $\Qc$-exact
change in the action, indicating that the correlators may be computed at weak
coupling, where the action localizes onto $\Qc$-invariant field configurations,
which in the case of the toric GLSM are given by the field configurations satisfying
\begin{eqnarray}
\label{eq:topconfig}
d \sigma_a & = & 0, \nonumber\\
\sum_a Q_i^a \sigma_a \phi^i & = & 0, \nonumber\\
D_a + f_a  & = & 0, \nonumber\\
D_{\bar{z}} \phi^i & = & 0
\end{eqnarray}
modulo the action of the gauge group. Morrison and Plesser have 
argued that this space of field configurations is a union of
disconnected moduli spaces  $\Mn$ labeled by the $u$ 
instanton numbers $n_a$.   Each $\Mn$ is a toric variety
which can be explicitly constructed as a holomorphic quotient.
Furthermore, they showed  that, fixing a phase with cone $\K$, 
the correlation functions 
\begin{equation}
Y_{\As} = \langle \sigma_{a_1}(z_1)\cdots\sigma_{a_s}(z_s) \rangle
\end{equation}
may be computed as a sum over $\nv$ in the dual cone $\Kd$:
\begin{equation}
\label{eq:ysum}
Y_{\As} = \sum_{\nv \in \Kd} Y_{\As}^{\nv} \prod_a q_a^{n_a},
\end{equation}
where $Y_{\As}^{\nv}$ is a coefficient independent of $\qv$, and the factor
$\prod_a q_a^{n_a}$ corresponds to 
$\exp(-S)$ evaluated on a field configuration belonging to $\Mn$.  Recall that 
$\Kd$ is defined by
\begin{equation}
\Kd = \left\{ \nv \in \left(\R^u\right)^\vee | \la \nv, \rv \ra \ge 0 
       ~~\text{for all}~~ \rv \in \K\right\}.
\end{equation}
{}From the form of $\Kd$ it follows that the sum in eqn. (\ref{eq:ysum}) 
converges for $\qv$ corresponding 
to $\rv$ in the interior of $\K$, with the radius of convergence determined 
by the distance to the closest 
singularity.  The coefficients $Y_{\As}^{\nv}$ are determined as 
intersection numbers on the toric variety $\Mn$.
These correlators are subject to conservation of ghost number, eqn. (\ref{eq:ghostnum}),
which implies 
\begin{equation}
\label{eq:ghnum}
Y_{\As}^{\nv} = 0 ~~~~\text{for} ~~~~ s \neq d + \sum_{a,i} n_a Q_i^a.
\end{equation}

We will not review the details of the computation of the non-zero $Y_{\As}^{\nv}$ in
this section, but we will highlight aspects of the 
computation that differ somewhat from the results of Morrison and Plesser.  Briefly, 
the method of \cite{MP:summing} is to recast the K\"ahler quotient of eqn. (\ref{eq:topconfig})
as a holomorphic quotient:
\begin{equation}
\label{eq:Mnform}
\Mn =  \frac{\bigoplus_i H^0(\O(d_i)) - F_{\nv} }{\left(\C^*\right)^u},
\end{equation}
where $d_i = \la \nv, \Qv_i\ra = \sum_a n_a Q_i^a$, 
$\O(n)$ is a holomorphic line bundle of degree $n$
over $\P^1$, $H^0(\O)$ is the space of holomorphic sections of $\O$, and $F_{\nv}$
is a certain excluded set, presented as an irreducible union of intersections of 
hyperplanes.

The $Y_{\As}^{\nv}$ are given by
\begin{equation}
\label{eq:yn}
Y_{\As}^{\nv} = f(\mu)\frac{1}{|H|} 
\langle \hat{\sigma}^{(\nv)}_{a_1} \cdots \hat{\sigma}^{(\nv)}_{a_s} \ECn \rangle_{\Mn},
\end{equation}
where $f(\mu)$ is a factor associated to twisting the theory, $H$ is (the possibly trivial) 
discrete subgroup of the gauge group left unbroken in the phase $\K$, the 
$\hat{\sigma}^{(\nv)}_a$ are proportional to toric divisors 
$\sigma^{(\nv)}_a$ on $\Mn$, and $\ECn$ is the Euler class of a certain obstruction 
bundle, whose appearance is described in \cite{MP:summing,AM:tqft}.  The intersection 
computations can be performed with standard toric methods.\footnote{More details on the 
geometry of $\Mn$ and the form of $\ECn$ may be found in Appendix \ref{app:toric}.}
 
The form of $f(\mu)$ and the relation between $\sigma^{(\nv)}$ and $\hat{\sigma}^{(\nv)}$
could be fixed by carrying out the path integral localization explicitly.  However, to
date, this has not been done.  Nevertheless, we can determine the correct scaling by 
appealing to the simple fact that the topological correlators are RG invariants, and thus 
should not depend on $\mu$ or $q_1(\mu)$ separately, but only through the RG-invariant
combination $\mu^\Delta q_1(\mu)$.  A natural guess that satisfies this criterion is
$\hat{\sigma}^{(\nv)} = \mu \sigma^{(\nv)}$ and $f(\mu) = \mu^{-d}$.  
This scaling is a difference from \cite{MP:summing} that we would like to emphasize.  
In the case of \cite{MP:summing}, where the primary interest 
concerned computations in Calabi-Yau (i.e. superconformal) GLSMs, this scaling
did not play an important role, since the ghost number selection rule restricted non-zero
correlators to $s=d$, and $q_1$ was a parameter of the model. It would be nice to determine
$f(\mu)$ and the scaling of $\sigma^{(\nv)}$ by an explicit computation.

Another important point is that $\M_{\nv=0} \simeq V$, the
toric variety corresponding to $\K$.  We will be interested in non-compact $V$,
for which the intersection computations are not well-defined, and we will need to
rely on another method to determine the zero instanton contributions.  

The most important difference from the superconformal case concerns the validity
of the instanton expansion in various phases. While in the superconformal case 
{\em any} phase could be used to compute the topological correlators via the instanton 
sum, in general, the instanton computation is only appropriate in certain phases.  When 
we compute in the high energy phase of the GLSM, where the Higgs branch is a good 
description of the semi-classical vacua, the instanton expansion is valid.  However, as 
we learned above, the low energy phases also possess a Coulomb branch, and an instanton 
expansion in that phase will inevitably miss these vacua.  
%%%%%%%%%%%%%%%%%%%%%%%%%%%%%%%%%%%%%%%%%%%%%%%%%%%%%%%%
\section{An example: {\protect $\Orbt$} } \label{s:Example}
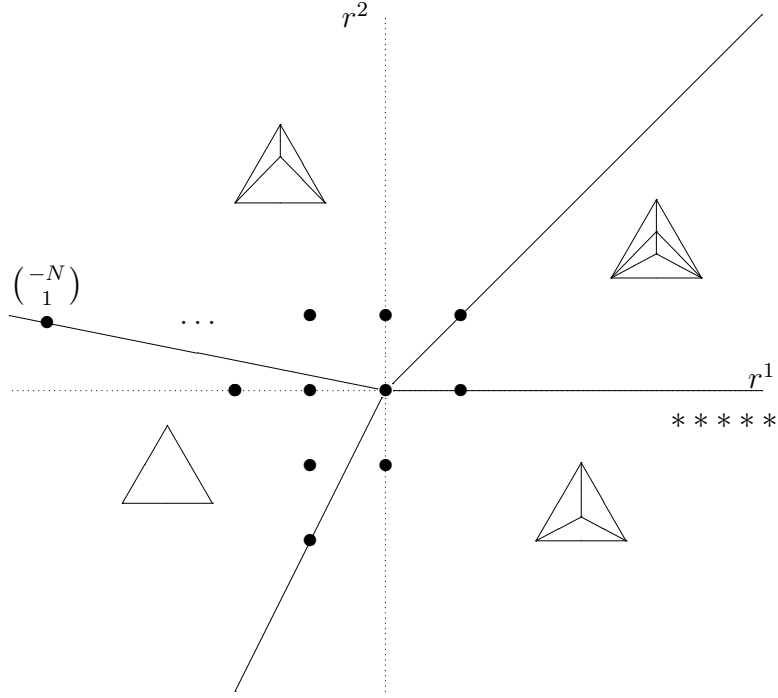
\begin{figure}
\[
\begin{xy} <1.0mm,0mm>:
  (0,0)*{\bullet} ="o", (50,0)*{}="1", (50,50)*{}="2", (-50,10)*{}="3", 
  (-20,-40)*{}="4", (-50,0)*{}="7", (10,0)*{\bullet}, (0,10)*{\bullet}, 
  (-10,-20)*{\bullet}, (10,10)*{\bullet}, (-10,10)*{\bullet}, 
  (-20,0)*{\bullet}, (-20,0)*{\bullet}, (-10,0)*{\bullet}, 
  (0,-10)*{\bullet}, (-10,-10)*{\bullet}, (-45, 9)*{\bullet},
  (0,50)*{}="5", (0,-40)*{}="6", (50,2)*{r^1}, (-4,50)*{r^2}, (-25,9)*{\ldots},
  (-45,14)*{\left(\genfrac{}{}{0pt}{}{-N}{1}\right)}, (45,-4)*{*****}="10",
  (30,15)*\xybox{ <0.3mm,0mm>:
  (0,0)*{}="1",(40,0)*{}="2",(20,34.64)*{}="3", (20,10.66)*{}="4", (20,20.43)*{}="5",
  \ar@{-}|{} "1";"2" \ar@{-}|{} "2";"3"
  \ar@{-}|{} "3";"1" \ar@{-}|{} "3";"5"
  \ar@{-}|{} "5";"4" \ar@{-}|{} "4";"1"
  \ar@{-}|{} "4";"2" \ar@{-}|{} "5";"1"
  \ar@{-}|{} "5";"2" },
  (-20,25)*\xybox{ <0.3mm,0mm>:
  (0,0)*{}="1",(40,0)*{}="2",(20,34.64)*{}="3", (20,20.43)*{}="5",
  \ar@{-}|{} "1";"2" \ar@{-}|{} "2";"3"
  \ar@{-}|{} "3";"1" \ar@{-}|{} "5";"1"
  \ar@{-}|{} "5";"2" \ar@{-}|{} "5";"3"
  },
  (20,-20)*\xybox{ <0.3mm,0mm>:
  (0,0)*{}="1",(40,0)*{}="2",(20,34.64)*{}="3", (20,10.66)*{}="4",
  \ar@{-}|{} "1";"2" \ar@{-}|{} "2";"3"
  \ar@{-}|{} "3";"1" \ar@{-}|{} "4";"1"
  \ar@{-}|{} "4";"2" \ar@{-}|{} "4";"3"
  },
  (-35,-15)*\xybox{ <0.3mm,0mm>:
  (0,0)*{}="1",(40,0)*{}="2",(20,34.64)*{}="3"
  \ar@{-}|{} "1";"2" \ar@{-}|{} "2";"3"
  \ar@{-}|{} "3";"1"
  },
\ar@{-}|{} "o"; "1" \ar@{-}|{} "o"; "2"
\ar@{-}|{} "o"; "3" \ar@{-}|{} "o"; "4"
\ar@{.}|{} "5"; "6" \ar@{.}|{} "7"; "1"
\end{xy}
\]
\caption{Phases of {\protect $\Orbt$}. The $*****$ indicates the projection of
         the semi-classical singularity to the $(r^1,r^2)$-plane.}
\label{fig:phases}
\end{figure}

Having reviewed the basic physics of the toric GLSM and its $A$-model,
we will now apply these ideas to a concrete example.  We build the GLSM 
corresponding to $\Orbt$ by starting with a minimal resolution of 
this toric singularity.  The toric construction of these orbifolds is well
explained in \cite{Craw:phd}. The toric fan of this minimal resolution 
has its one-dimensional rays labeled by
\begin{equation}
\left( \begin{array}{c} v_1 \\ v_2 \\ v_3 \\ v_4  \\v_5\end{array} \right)=
\left( \begin{array}{ccc}
         M    & 0    & 0 \\
         0    & M    & 0 \\
         0    & 0    & M \\
         2    & 2    & 1 \\
         1    & 1    & N+1
       \end{array}
\right),
\end{equation}
where $M = 2N+1$.  The combinatorics of the fan are as follows:
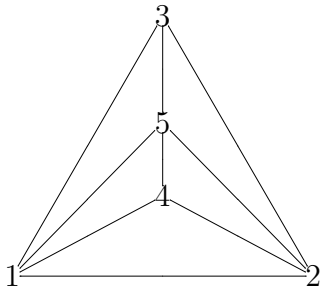
\begin{equation*}
\begin{xy} <1.0mm,0mm>:
  (0,0)*{1}="1",(40,0)*{2}="2",(20,34.64)*{3}="3", (20,10.66)*{4}="4", (20,20.43)*{5}="5",
  \ar@{-}|{} "1";"2"
  \ar@{-}|{} "2";"3"
  \ar@{-}|{} "3";"1"
  \ar@{-}|{} "3";"5"
  \ar@{-}|{} "5";"4"
  \ar@{-}|{} "4";"1"
  \ar@{-}|{} "4";"2"
  \ar@{-}|{} "5";"1"
  \ar@{-}|{} "5";"2"
\end{xy}
\end{equation*}
The GLSM gauge charges are a basis for the relations among the $v_i$.
A convenient choice is
\begin{equation}
 Q = \left(\begin{array}{ccccc}
                 1 & 1 & 1 &  -N & -1 \\
                 0 & 0 & 1 &   1 & -2
               \end{array}
         \right).
\end{equation}
There are four classical phases, each corresponding to a triangulation
of the fan.  The fan without any subdivisions is the orbifold phase, the
partially subdivided fans correspond to partial resolutions, and the completely
subdivided fan is the smooth phase.  These phases are depicted in Fig. (\ref{fig:phases}).
The RG running of this GLSM is determined by $\Delta = \sum_i Q_i^1 = 2 -N$:
for $N>2$ the model driven to small $|q_1|$ or, equivalently, large $r^1$. 
The figure encapsulates much of the the basic physics discussed above:  the
UV fixed point is an orbifold, and RG flow leads to a less singular geometry;
when $|q_1|$ is small and $q_2 =1/4$, semi-classical analysis shows that 
a continuous Coulomb branch, parametrized by the expectation value 
of $\sigma_2$ emerges. 

\subsection{Topological correlators}
The topological correlators of the model are 
$Y_{ab} = \langle \sigma_1^a \sigma_2^b \rangle$.  The ghost number anomaly implies 
that $Y_{a,b} = 0$ if $a+b \neq 3 + (2-N) n_1$ for some
$n_1$.  This immediately leads to a puzzle:  while expanding in the
instantons in the orbifold phase, where $\nv \in \Kd$ may have arbitrarily 
{\em negative} $n_1$, leads to an infinite number of non-trivial correlators, 
the instantons in the smooth phase, where $\nv \in \Kd$ requires $n_1 \ge 0$,  
cannot support most of these.  Based on the experience with Calabi-Yau 
GLSMs, this would be confusing,  because these correlation functions
are RG invariants, and should be computable in any phase.  Furthermore, since 
no singularities separate the
phases from one another, we expect to be able to analytically continue
the correlation functions from any one phase to another.  In fact, it
is known that the $Y_{a,b}$ are rational functions in the $q_{1,2}$,
so that this analytic continuation is unambiguous.  

Of course, there is a natural resolution to this puzzle:  the presence of
isolated Coulomb vacua deep in the interior of a phase, precisely where 
an instanton computation is expected to be trustworthy, means that the 
instanton computations may not be reliable in these phases.  The Coulomb
vacua  may not only invalidate a Higgs branch computation but also 
support the missing correlators.

\subsubsection{The instanton sums}
Using the methods of \cite{MP:summing}, we are able to explicitly
compute the non-zero correlators in the orbifold phase:
\begin{equation}
\label{eq:expcorr}
Y_{a,b} = q_1(\mu)^{n_1} 
\sum_{n_2 = N n_1}^{ \lfloor -\ff{n_1}{2} \rfloor} 
     q_2^{n_2} Y_{a,b}^{n_1,n_2}, ~~~ \text{with} ~~~ a+b = 3 +(2-N) n_1.
\end{equation}
The notation $\lfloor a \rfloor$ ($\lceil a \rceil$) signifies the
greatest (smallest) integer less (greater) than $a$ and the coefficients are
given by 
\begin{eqnarray}
\label{eq:yexmpl}
Y_{a,b}^{n_1,n_2}& =& \mu^{n_1 (2-N)} \ff{1}{M} \oint\limits_{C(0)} \frac{dz}{2\pi i} 
                    \frac{ (2+z)} {z (1 +(N+1)z)}
                    \left[ \frac{ (-M)^{N+1} z }{ (2+z)^N (1+(N+1)z)}\right]^{n_1}
                    \times\nonumber\\
                 &~& ~~~~~~~~~ \times\left[ \frac{-M z^2}{1+(N+1)z} \right]^{n_2}
                    \left[ \frac{-1+Nz}{2+z} \right]^b,
\end{eqnarray}
where, as before, $M=2N+1$ and$C(0)$ is a contour about $z=0$ that avoids all other poles
of the integrand.\footnote{Details of the computation may be found in 
appendix \ref{app:instsum}.} As mentioned above, the methods of \cite{MP:summing} are
reliable only for $n_1 \neq 0$.  The finite sum on $n_2$ implies that these 
correlators can only be singular as $q_2 \to \infty$ or $q_2 \to 0$.  On the other
hand, the instanton sums for the $n_1=0$ correlators seem to have just the $\nv =0$
term, and hence should just be constants.  So, the 
semi-classical singularity at $q_2 = \ff{1}{4}$ is another puzzle, since there 
do not seem to be correlators sensitive to it.  

In order to compute the $n_1 = 0$ correlators, as well as to make sense
of the puzzle of missing correlators, we can appeal to a remarkable
set of identities satisfied by the correlators.  Let $\O$ be an
$A$-model observable of this GLSM.  Then, using  eqn. (\ref{eq:yexmpl}), 
it is easy to show that the $n_1\neq 0$ correlators satisfy
\begin{eqnarray}
\label{eq:mrexample}
 \langle \left(\sigma_1 + \sigma_2\right) 
         \left(\sigma_2 -N \sigma_1\right) \O \rangle & = & 
  q_2 \langle \left(\sigma_1 + 2 \sigma_2\right)^2 \O \rangle, \nonumber\\
 \langle \left(\sigma_1^3 + \sigma_1^2\sigma_2\right) \O \rangle & = &
 -\mu^{(2-N)} q_1(\mu) \langle \left(\sigma_2 -N\sigma_1\right)^N \left(\sigma_1 + 2\sigma_2\right) \O
 \rangle.
\end{eqnarray}
We will show below that these relations hold more generally and are
quantum cohomology relations on topological correlators, whose form is
determined by the effective twisted superpotential.  Assuming  that 
these relations are satisfied by the $n_1=0$ correlators
as well as the $n_1\neq 0$ ones, it is straightforward to use the relations to determine 
the $Y_{3-b,b}$ correlators:  
\begin{eqnarray}
Y_{3,0} & = &  \frac{2}{M} \nonumber,\\
Y_{2,1} & = & -\frac{1}{M} \nonumber,\\
Y_{1,2} & = &  \frac{ 2 q_2 -N-1 }{M (4q_2 -1)} \nonumber,\\
Y_{0,3} & = &  \frac{-4q_2^2 + (5+6N)q_2 + N^2-N-1}{M\left(4q_2-1\right)^2}.
\end{eqnarray}
Thus, we find that the $Y_{1,2}$ and $Y_{0,3}$ are the correlators sensitive 
to the $q_2 = \ff{1}{4}$ singularity.  We stress that this $q_2$ dependence
means that these correlators are not computed by the standard instanton sums.

\subsection{ Relations and missing correlators  }
The relations in eqn. (\ref{eq:mrexample}) are powerful.  We can easily
show that they completely determine all the instanton correlators in terms
of {\em two} of them, say $Y_{3,0}$ and $Y_{2,1}$.  This may well be
the most efficient procedure for evaluating $A$ model correlators in toric
GLSMs, and we will return to it below.  However, they are also interesting
for our present purpose of understanding the ``disappearing'' of the Higgs
phase correlators.

There is a simple way to write down a set of correlators that obeys
eqn. (\ref{eq:mrexample}).  It is  based on the isolated Coulomb vacua 
of the GLSM.  The equations of motion that follow from the twisted superpotential
of this model are (compare to eqn. (\ref{eq:mrexample}) )
\begin{eqnarray}
\left(\sigma_1 + \sigma_2\right) \left(\sigma_2 -N \sigma_1\right)  & = & 
  q_2 \left(\sigma_1 + 2 \sigma_2\right)^2 , \nonumber\\
\left(\sigma_1^3 + \sigma_1^2\sigma_2\right) & = &
 -\mu^{(2-N)} q_1(\mu) \left(\sigma_2 -N\sigma_1\right)^N \left(\sigma_1 + 2\sigma_2\right).
\end{eqnarray}
As discussed in section \ref{ss:coulombvac}, there are $N_v = 2 (N-2)$
isolated vacua.  Let us label these by $(\sigma_{1,\alpha}, \sigma_{2,\alpha})$,
$\alpha = 1,\ldots, N_v$.  A set of functions that obey the 
relations and  are rational functions of the $q_a$ is given by
\begin{equation}
\label{eq:coulcorrex}
Y_{a,b}^{\text{Coul}} = \sum_{\alpha=1}^{N_v}
                        Z(\sigma_{1,\alpha}, \sigma_{2,\alpha}) 
                        \sigma_{1,\alpha}^a \sigma_{2,\alpha}^b,
\end{equation}
with $Z(\sigma_1,\sigma_2)$ an undetermined function.  Recall that
the $(\sigma_{1,\alpha},\sigma_{2,\alpha})$ can be parametrized as
in eqn. (\ref{eq:teomsols}).  In this case, we find 
\begin{eqnarray}
P(\omega_\pm) & = & (1+\omega_\pm) (\omega_\pm-N) - q_2 (1+2\omega_\pm)^2 = 0,
                 \nonumber\\
\sigma_{1,\pm;p} & = & \lambda^p \mu \left(q_1(\mu) s(\omega_\pm)\right)^{\frac{1}{2-N}}, \nonumber\\
\sigma_{2,\pm;p} & = & \omega_\pm \sigma_{1,\pm;p},
\end{eqnarray}
where $\lambda = e^{\frac{2 \pi i}{ N-2}}$,  $p=0,\ldots, N-1$, and
\begin{equation}
s(\omega) = \frac{(-1-2\omega) (\omega-N)^N}{1+\omega}.
\end{equation}
If, in addition to the relations, we demand that the selection rule
$a+b = 3 + (2-N) n_1$ holds for non-zero correlators, the function $Z$ is
constrained to have the form 
\begin{equation}
Z(\sigma_{1,\pm;p}, \sigma_{2,\pm;p}) = \sigma_{1,\pm;p}^{-3} F(\omega_\pm)/(N-2),
\end{equation}
for some $F(\omega)$.
The two instanton correlators that are left undetermined by the relations
are just right to fix the form of $F(\omega_\pm)$, and the correlators 
can be written as in eqn. (\ref{eq:coulcorrex}) with 
\begin{equation}
\label{eq:Zexmpl}
Z(\sigma_1,\sigma_2) = \frac{1}{(N-2)\sigma_1^2 (2 \sigma_2 N + (3N+1)\sigma_1) }.
\end{equation} 
%Note, this $Z$ is determined from {\em minus} the instanton sums computed in
%the orbifold phase (eqn. (\ref{eq:yexmpl})), since, as discussed above, 
%we expect these to correspond to the $A$-model correlators.

The form of $Y_{a,b}$ that follows from $Y_{a,b} = Y_{a,b}^{\text{Coul}}$ 
is suggestive of a correlator
computation on the Coulomb branch.  Recall, that the Coulomb vacua are
massive, and with eqn. (\ref{eq:topconfig}) precluding tunneling between
the vacua ($d\sigma = 0$), it is sensible that the correlators
are given simply by products of the vacuum expectation values of the fields.
Furthermore, the appearance of $Z$ is also natural:  we expect that
the localization of the GLSM path integral in the $A$-model and the 
integrating-out of the chiral matter and the anticommuting
fields will lead to a non-trivial measure for the $\sigma_a$ path 
integral.

This then is the resolution of the puzzle:  the correlators that were
computed in the orbifold phase as an instanton expansion about a Higgs
vacuum do not disappear.  Instead, they become correlators supported
on the Coulomb vacua. 
%%%%%%%%%%%%%%%%%%%%%%%%%%%%%%%%%%%%%%%%%%%%%%%%%%%%%
\section{Quantum cohomology relations and the Coulomb branch} \label{s:MR}
In this section we study the relations of eqn. (\ref{eq:mrexample}) in a more
general setting.  The general form of these relations follows from 
eqn. (\ref{eq:teompoly}).  
Letting $\xi_i = \sum_a Q_i^a \sigma_a$, we have
\begin{equation}
\label{eq:mrexpl}
\la \O \prod_{Q^a_i > 0} \xi_i^{Q_i^a} \ra = 
\mu^\Delta_a q_a(\mu) \la \O \prod_{Q^a_i < 0} \xi_i^{-Q_i^a} \ra  ~~~\text{for all}~~a,
\end{equation}
where $\O$ is, an $A$-model observable.  These {\em quantum cohomology
relations} are proved in Appendix \ref{app:mrproof}.  The proof assumes
that there exists some phase $\K$ where the correlators may be computed by 
instanton sums.  Working in this phase, we are able to reduce eqn. (\ref{eq:mrexpl}) 
to statements about the cohomology rings of the moduli spaces $\Mn$.  Given
the toric structure of $\Mn$, we show that these statements follow from a
few simple combinatoric observations.

We can show that the requisite $\K$ exists in general GLSMs.  The phase $\K$ 
must be one of the UV phases of the  GLSM.  Working with a standard basis, we 
see that in a UV phase $r^1 = |r^1|\sign(\Delta)$.  The classical $D$-term 
equations for $\rv \in \K$ take the form
\begin{eqnarray}
\sum_i Q^1_i |\phi^i|^2 & = & |r^1| \sign(\Delta),\nonumber\\
\sum_i Q^a_i |\phi^i|^2 & = & r^a, ~~\text{for}~a>1.
\end{eqnarray}
The first of these obviously has a solution for arbitrarily large $|r_1|$, and the 
others have the property that if there is no solution for $r^a$, then there exists 
a solution for $-r^a$.  Thus, there exists a UV phase $\K$ where, by
suitably tuning the F-I parameters, the classical Higgs branch can be made to be 
an arbitrarily good description of the physics, and thus, an instanton computation
should be reliable.

As in the example, it is easy to see that these relations determine
the correlators up to a finite subset.  Strictly speaking, the 
instanton correlators $\langle \prod_{a=1}^u \sigma_a^{A_a}\rangle$ 
are only defined for $A_a \ge 0$ and $\sum_{a} A_a \ge d$, and if 
one attempts to extend the $A_a$ outside the physical range by using the 
quantum cohomology relations, it is possible that one will run into
inconsistencies.  However, when the GLSM has, in addition to a pure
Higgs phase, a phase with isolated Coulomb vacua, we believe the
relations can be extended consistently to generate a full $\Z^u$ lattice
of functions which obey the quantum cohomology relations and the ghost
number selection rule.  We will call this property {\em extendability}.
We will show below that this property holds in a wide class of examples. 

Extendability is a necessary condition for the following conjecture: 
the $A$-model correlators of a toric GLSM with a pure Higgs phase and 
a phase with isolated Coulomb vacua can be put into the general form
\begin{equation}
\label{eq:standardform}
Y_{A_1,\ldots,A_u} = \langle \prod_{a=1}^u \sigma_a^{A_a}\rangle = 
\sum_{\alpha=1}^{N_v} Z(\sigma_{1,\alpha}, \ldots,\sigma_{u,\alpha})
                                  \prod_a \sigma_{a,\alpha}^{A_a},
\end{equation}
where $(\sigma_{1,\alpha},\ldots,\sigma_{u,\alpha})$, $\alpha=1,\ldots,N_v$ are
the isolated Coulomb vacua.  

It should be possible to prove this in all generality.  However, as we 
understand it now, the proof is likely to require some more general tools
of Gr\"obner  bases and elimination theory.  We will 
content ourselves with a proof of the conjecture for the case $u=2$.  

Consider a toric GLSM with $u=2$ which satisfies the assumptions above and,
in addition, has an irreducible $P(\omega)$.
The presence of the pure Higgs phase allows us to assert that the $A$-model
correlators are rational functions of $q_1,q_2$, satisfying eqn. (\ref{eq:mrexpl}) 
and obeying the selection rule that $Y_{A,B} = 0$ unless $A+B = d  + \Delta n_1$.
Since the GLSM also has a phase with isolated Coulomb vacua, it follows that a
 candidate set of correlators that obey the relations is given in 
eqn. (\ref{eq:standardform}), with the $N_v = \deg(P) |\Delta|$ vacua 
determined by the twisted equations of motion.  If the GLSM correlators are
extendable to $(A,B) \in \Z^2$, then it is possible that there exists a 
$Z(\sigma_1,\sigma_2)$ such that the correlators are computed by eqn. (\ref{eq:standardform}).
We will now construct the required $Z(\sigma_1,\sigma_2)$.

The selection rule $A+B = d+\Delta n_1$ implies that
\begin{equation}
Z(\sigma_1,\sigma_2) = \frac{1}{|\Delta|} \sigma_1^{-d} F(\omega),
\end{equation}
in which case eqn. (\ref{eq:standardform}) reduces to 
\begin{equation}
\label{eq:fstandard}
Y_{d+\Delta n_1 -B, B} = \left(\mu^\Delta q_1\right)^{n_1} 
\sum_{\omega | P(\omega) = 0} s(\omega)^{n_1} \omega^B F(\omega).
\end{equation}
If we label the roots of $P(\omega)$ by $\omega_s$, $s=1,\ldots,\deg(P),$ then
it is straightforward to determine $F_s = F(\omega_s)$ in terms of the 
$Y_{d-B,B}$:
we merely need to solve the linear system
\begin{equation}
\sum_{s} M_{ts} F_s = Y_{d-t,t},
\end{equation}
where $M_{ts} = (\omega_s)^{t}$ is the Vandermonde matrix for the polynomial
$P(\omega)$.  $M$ is invertible provided that the discriminant of $P$ is non-zero.
This will be true for generic $q_2$.  Having determined the $F_s$, we can 
construct an $F(\omega)$:
\begin{equation}
F(\omega) = \sum_{s=1}^{\deg(P)} 
\frac{P(\omega) F_s}{(\omega-\omega_s) P'(\omega_s)}.
\end{equation}
Obviously, $F(\omega)$ is {\em not} determined uniquely:  the
correlators are unchanged under the shift 
$F(\omega) \to F(\omega) + P(\omega) g(\omega)$, where
$g(\omega)$ is any function that is non-singular at the roots of $P(\omega)$.

This elegant form depends on the assumption that $P(\omega)$
is irreducible.  This assumption was made
in a slightly more general form in writing a polynomial form for the 
equations of motion (eqn. (\ref{eq:teompoly})).  It seems that the
proper generalization is to work with $\hat{P}(\omega)$, the irreducible,
$q_2$ dependent factor of $P(\omega)$, as the roots of $\hat{P}(\omega)$ will
actually correspond to the $\sigma$-vacua.  It is conceivable that in the
case that $P(\omega)$ is reducible, one may be able to write down a
{\em stronger} set of quantum cohomology relations than the ones in 
eqn. (\ref{eq:mrexpl}).  However, these stronger relations are not needed for any 
of the arguments above.  Thus, we expect that the reduction to $\hat{P}(\omega)$
is all that is necessary to handle the case of reducible $P(\omega)$.\footnote{
This is the description for $u=2$.  Its generalization to $u>2$ is obvious.}

Although our approach uses the quantum cohomology relations to argue that $A$-model 
correlators are computed by the Coulomb branch in the IR phases, and we cannot 
explicitly perform the Coulomb branch computation, it is interesting to consider the 
converse argument: if the $A$-model correlators are computed by the 
Coulomb branch, then they obey the quantum cohomology relations.  If true, it 
would provide a physical explanation for the quantum cohomology relations, which 
are very obscure from the point of view of the toric geometry of the $\Mn$.
%%%%%%%%%%%%%%%%%%%%%%%%%%%%%%%%%%%%%%%%%%%%%%%%%%%%%%%%%%%%%%%%%
\section{Two-parameter models} \label{s:General}
The reader may wonder to what extent the physics of the example of the $\Orbt$ 
GLSM resembles the general GLSM.  In this section we will show that some
generalizations are straightforward and enlightening.  We stick to two-parameter
models because of their simplicity.
\subsection{General orbifold GLSMs with $u=2$}
The simplest generalization is to 
consider the most general two-parameter model with a $\Orb$ phase.  
The toric fan of the minimal model for these GLSMs is easily written 
down.  It has the same combinatorics  as the toric fan 
of $\Orbt$, except that the one dimensional
cones are given by\footnote{We do not prove this, but
it can probably be proved by using the results of Seb\"o on
Hilbert bases\cite{Sebo:hilb}.}
\begin{equation}
\left( \begin{array}{c} v_1 \\ v_2 \\ v_3 \\ v_4  \\v_5\end{array} \right)=
\left( \begin{array}{ccc}
         km-1    & 0       & 0    \\
         0       & km-1    & 0    \\
         0       & 0       & km-1 \\
         k       & k       & 1 \\
         1       & 1       & m 
       \end{array}
\right),
\end{equation}
where $k,m$ are positive integers and $km > 1$.
The charges that follow from this fan are 
\begin{equation}
\label{eq:Qgen}
Q = \left( \begin{array}{ccccc}
            0 & 0 & 1 &  1  & -k \\
            1 & 1 & 0 & -m  &  1
           \end{array}
    \right).
\end{equation}
This explicit form of $Q$ is useful.  For example, the results
of the previous section relied on $P(\omega)$ being an irreducible
polynomial.   Due to the simple form of $P(\omega)$ 
(eqn. (\ref{eq:teomsols})), it is easy to state this as a condition
on the $\Qv_i$:  if $Q_i^2 \neq 0$, then there does not exist
$\alpha \in \Q_{>0}$ such that $\Qv_i = - \alpha \Qv_j$ for
some $j$.   The existence of such an $\alpha$ is, of course,
independent of the choice of basis for the $\Qv_i$. 
Thus, from eqn. (\ref{eq:Qgen}), it follows that $P(\omega)$ has
no $q_2$-independent roots, and the results of section \ref{s:MR}
apply to these models directly.

Applying the standard techniques outlined in section \ref{s:Example},
we can see that the phase diagram is similar to the example
we have studied.  Even without explicitly studying the correlators 
in these models, it is clear that their general structure would 
follow that of the example.  Nevertheless, studying
this set of models in greater detail may be a way towards 
understanding the form of $Z$ in eqn. (\ref{eq:standardform}) 
directly from the combinatorial data of the GLSM.  This would enable us to
write down the $A$-model correlators without ever doing an
instanton sum.  We will realize this hope, at least for a large
class of GLSMs, in the next section. 

\subsection{Topological correlators in orbifold models}
Suppose we are given a GLSM with a phase where the $A$-model 
correlators may be computed by instanton sums.  If we were able
to write down a general expression like eqn. (\ref{eq:expcorr}) 
for a (large enough) subset of these correlators, then, since
these satisfy the quantum cohomology relations, we could 
use the results of section \ref{s:MR} to extract the 
$Z(\sigma_1,\sigma_2)$ or, equivalently, $F(\omega)$.  This is
a formidable task, since for arbitrary GLSMs, even with
$u=2$, the intersection theory of $\Mn$ can be quite complicated.
However, there is one situation where the $\Mn$ have a simple structure---
when the  Higgs phase $\K$ 
corresponds to an orbifold.   In fact, a few more assumptions
are needed to make the computation tractable, but with these
in hand, we will be able to extract $F(\omega)$.

\begin{figure}
\[
\begin{xy} <1.0mm,0mm>:
  (0,0)*{\bullet} ="0", (12,0)*{}="1", (0,40)*{}="2", (-50,25)*{}="3", 
  (-20,-40)*{}="4", (-50,0)*{}="5", (0,-50)*{}="6", 
  (10,-4)*{r_1}, (-4,38)*{r_2}, (-25,12.5)*{\bullet}, 
  (-19,15)*{\frac{1}{g_4} \Qv_4}, 
  (-15, -30)*{\bullet}, (-10,-30)*{\frac{1}{g_5}\Qv_5},
  (-45,-10)*{\K},
  (-35,-15)*\xybox{ <0.3mm,0mm>:
  (0,0)*{}="1",(40,0)*{}="2",(20,34.64)*{}="3",
  \ar@{-}|{} "1";"2" 
  \ar@{-}|{} "2";"3"
  \ar@{-}|{} "3";"1"
  },
\ar@{-}|{} "0"; "3"
\ar@{-}|{} "0"; "4"
\ar@{.}|{} "1"; "5" 
\ar@{.}|{} "2"; "6" 
\end{xy}
\]
\caption{The orbifold phase in a general two-parameter model. 
         {\protect $g_{4,5} = \gcd(Q^1_{4,5},Q^2_{4,5})$}. }
\label{fig:gphase}
\end{figure}
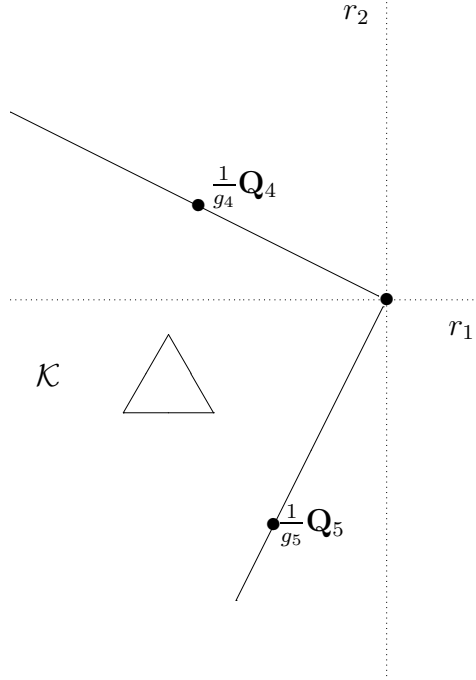

Suppose we are given a GLSM with charges $Q_i^a$, $i=1,\ldots,5$ 
(generalization to $i=1,\ldots,n$ is straightforward), and $a=1,2$.
Suppose that, in addition, the GLSM satisfies the following properties:
\begin{itemize}
\item[-] The $Q_i^a$ are in a standard basis, with $\Delta = \sum_i Q_i^1 <0$;
\item[-] the phase diagram has a phase $\K$ as shown in Figure (\ref{fig:gphase}).
\end{itemize}
From these assumptions it immediately follows that 
for any non-zero $\nv \in \Kd$, $d_i =\la \nv, \Qv_i \ra $ satisfy
 \begin{equation}
  \label{eq:genpropdi}
  d_i < 0   ~\text{for}~   i \le 3, ~~~~ d_i \ge 0 ~\text{for}~   i = 4,5.
 \end{equation} 
With these assumptions, the non-zero correlators take the form
\begin{equation}
Y_{3+\Delta n_1 - b, b} = \left(\mu^\Delta q_1(\mu)\right)^{n_1} 
                          \sum_{n_2 = N_-}^{N_+} q_2^{n_2} Y_{3+\Delta n_1-b,b}^{n_1,n_2},
\end{equation}
where $N_+ = \lfloor -\frac{Q_5^1}{Q_5^2} n_1 \rfloor$, and 
      $N_- = \lceil-\frac{Q_4^1}{Q_4^2}n_1 \rceil$.\footnote{ Recall that $\lfloor a \rfloor$ ($\lceil a \rceil$)
is the greatest (smallest) integer less (greater) than $a$.}
The coefficients $Y_{3+\Delta n_1 -b,b}^{n_1,n_2}$ are determined  
by a tractable intersection computation on $\Mn \simeq \P^{d_4}\times\P^{d_5}$.  
Omitting some straightforward steps (illustrated for the example of $\Orbt$
in appendix \ref{app:instsum}), we find that for $n_1 <0$ correlators,
\begin{equation}
\label{eq:ygen}
Y_{3+\Delta n_1-b,b} = (\mu^\Delta q_1)^{n_1} 
                   \oint\limits_{C\left(\omega_\ast\right)} \frac{d\omega}{2\pi i}
 \omega^b \prod_i \zeta_i^{-Q_i^1 n_1 -1} \prod_{Q_i^2 > 0} \zeta_i^{Q_i^2}
 \left( q_2 \prod_i \zeta_i^{-Q_i^2} \right)^{N_-} P(\omega)^{-1},
\end{equation}
where $\omega_\ast = -Q^1_5/Q^2_5$.
The integrand has, in addition to the pole at 
$\omega = -Q^1_5/Q^2_5$, a set of poles at the roots of $P(\omega)$ 
and a set of potential $q_2$-independent poles at 
\begin{eqnarray*}
 \omega & = & \infty, \\
 \zeta_i & = & 0, ~~~i \le 4. \\
\end{eqnarray*}
We will now show that for $n_1 <0$ these potential poles are not poles at all.
First, for large $\omega$, the integrand scales like $\omega^{A_\infty-1} d\omega$, with
\begin{equation}
A_\infty  = 1 + b - \deg(P) + \sum_{i|Q_i^2 > 0} Q_i^2 + \sum_{i|Q^2_i\neq 0} \left( -Q_i^1 n_1  -1 -Q_i^2 N_-\right).
\end{equation}
Since $\deg(P) = \sum_{i|Q_i^2 >0} Q_i^2$ and $0 \le b \le 3+n_1 \Delta$, 
\begin{equation}
A_\infty \le -1 + \sum_{i|Q^2_i=0} (Q^1_i n_1 +1).
\end{equation}
But, 
\begin{equation}
\sum_{i|Q^2_i = 0 } ( Q^1_i n_1 + 1) = \sum_{i|Q^2_i=0, i<4} ( \la (n_1,0), \Qv_i \ra + 1) \le 0,
\end{equation}
and since $ (n_1,0) \in \Kd$, it follows from eqn. (\ref{eq:genpropdi}) that $A_\infty \le -1$, 
and there is no pole at $\omega = \infty$.
Next, near $\zeta_i =0$, $i\le 3$, the integrand scales as $\zeta_i^{A_i}$, with 
\begin{equation}
A_i \ge -Q_i^1 n_1 -1 -Q_i^2 N_- = -1 - \la \left(n_1,N_-\right), \Qv_i\ra,
\end{equation}
and since $\left(n_1,N_-\right) \in \Kd$, eqn. (\ref{eq:genpropdi}) guarantees $A_i > -1$. 
Finally, consider the integrand near $\zeta_4 = 0$.  It scales as $\zeta_4^{A_4}$ with
\begin{equation}
A_4 = -Q^1_4 n_1 -1 + Q^2_4 - Q^2_4 \lceil -\frac{Q^1_4}{Q^2_4} n_1 \rceil \ge Q^2_4 -1 > -1,
\end{equation}
since $Q^2_4 > 0$ by assumption.
It follows that we may write the correlators as\footnote{We have left off the factor of
$\left(q_2 \prod_i \zeta_i^{-Q_i^2} \right)^{N_-}$, since at $\omega = \hat{\omega}$ it
is $1$.}
\begin{equation}
\label{eq:corans}
Y_{3+\Delta n_1 -b,b}  =  - \left(\mu^\Delta q_1\right)^{n_1} 
                   \sum_{\hat{\omega} | P(\hat{\omega}) =0} ~\oint\limits_{C(\hat{\omega})} 
                   \frac{d\omega}{2\pi i} s(\omega)^{n_1} \omega^b 
                   \frac{\prod_{i|Q_i^2 > 0} \zeta_i^{Q_i^2}}{P(\omega) \prod_i \zeta_i},
\end{equation}
where, as before, $s(\omega) = \prod_i \zeta_i^{-Q^1_i}$.  Thus, we have an explicit
demonstration that the $n_1<0$ correlators in these models satisfy the quantum cohomology
relations, and the relations are extendable to all $a,b$ that satisfy $a+b = 3+ \Delta n_1$.
This means the correlators may be written in the form of eqn. (\ref{eq:fstandard}).  It
is important to note that the $n_1=0$ correlators are not computed correctly by 
eqn. (\ref{eq:ygen}) with $n_1=0$, as, in particular, such expressions would not satisfy
the quantum cohomology relations.  However, they are correctly computed by  eqn. (\ref{eq:corans}).
The difference between the two expression can be traced to the appearance of extra poles in
the integrand of eqn. (\ref{eq:ygen}) when $n_1 = 0$.

We now wish to compare eqn. (\ref{eq:corans}) to eqn. (\ref{eq:fstandard}), which may be written
as
\begin{equation}
Y_{3+\Delta n_1 -b, b} = \left(\mu^\Delta q_1\right)^{n_1} 
\sum_{\hat{\omega} | P(\hat{\omega}) = 0} ~\oint\limits_{C(\hat{\omega})} \frac{d\omega}{2\pi i}
             s(\omega)^{n_1} \omega^b F(\omega) \frac{P'(\omega)}{P(\omega)}.
\end{equation}
We may now extract $F(\omega)$:
\begin{equation}
F(\omega) = - \frac{\prod_{Q^2_i>0} \zeta_i^{Q^2_i}}{P'(\omega) \prod_{i} \zeta_i}.
\end{equation}
It can be shown that when $P(\omega) =0$,
\begin{equation}
P'(\omega) = \prod_{i|Q_i^2 > 0} \zeta_i^{Q_i^2} \sum_i \left(Q_i^2\right)^2 \zeta_i^{-1},
\end{equation}
so, finally,\footnote{The capricious way of writing $-1$ as $\sign(\Delta)$ will become
apparent in the next section.} 
\begin{equation}
\label{eq:F}
F(\omega) = \frac{\sign(\Delta)}{\sum_i \left(Q_i^2\right)^2 \prod_{j\neq i} \zeta_j},
\end{equation}
and the corresponding $Z(\sigma_1, \sigma_2)$ is 
\begin{equation}
\label{eq:Z}
Z(\sigma_1,\sigma_2)^{-1}  =  \Delta \sigma_1^{3} 
             \left .\sum_i (Q^2_i)^2 \prod_{j\neq i} \zeta_j \right|_{\omega=\sigma_2/\sigma_1}.
%
%Z(\sigma_1, \sigma_2) = - \frac{1}{\sigma_1^{3-p} 
%                                         \sum_i \left(Q_i^2\right)^2 \prod_{j\neq i}
%                                                \left(Q_j^1 \sigma_1 + Q_j^2\sigma_2\right)}.
\end{equation}
 
\subsection{A few examples}
The example  of $\Orbt$ satisfies all the assumptions for our general treatment,
and it is easy to verify that eqn. (\ref{eq:Z}), specialized to this case is 
eqn. (\ref{eq:Zexmpl}).  What is more surprising is that eqns. (\ref{eq:F},\ref{eq:Z}) give 
the correct correlators in more general models.  We have studied the following cases 
in detail.
\begin{itemize}
 \item[-] The $\C^3/\Z_{3(211)}$ GLSM.  This model has charges
 \begin{equation}
  Q = \left(\begin{array}{ccccc}
                  0 & 1 & 1 &  1 & -2 \\
                  1 & 0 & 0 & -2 & 1
               \end{array}
         \right).
 \end{equation}
 Thus, $\Delta = 1 > 0$, and we compute the instanton sum in the
 smooth phase, where there exist sub-cones in $\Kd$---the $d_i$ change
 signs in $\Kd$.
 The explicit computation of the instantons yields
 \begin{equation}
  F(\omega) = -\frac{1}{2\left(1+\omega\right)}.
 \end{equation}
 \item[-] The $\C^3/\Z_{5(211)}$ GLSM.  This model has charges
 \begin{equation}
  Q = \left(\begin{array}{ccccc}
                  0 & 1 & 1 &  1 & -2 \\
                  1 & 1 & 1 & -2 & -1
               \end{array}
         \right).
 \end{equation}
 So, $\Delta = 1 >0$ again.  The model has a unique 
 UV phase---one of the partially resolved phases, and the
 instanton sum in this phase also has sub-cones.  The
 smooth and orbifold phases are intermediate (in the sense of the RG flow).
 When we perform the instanton sums,  we find
 \begin{equation}
  F(\omega) = -\frac{1}{2 \left(1+5 \omega +\omega^2\right)\left(1 +\omega\right)}.
 \end{equation}
 \item[-] Finally, consider a GLSM without a $\C^3/\Z_N$ orbifold phase.
 This model has the fan
 
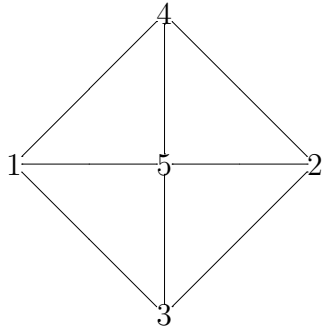
\begin{equation*}
 \begin{xy} <1.0mm,0mm>:
   (0,0)*{1}="1",(40,0)*{2}="2",(20,-20)*{3}="3", (20,20)*{4}="4", (20,0)*{5}="5",
   \ar@{-}|{} "1";"5"
   \ar@{-}|{} "2";"5"
   \ar@{-}|{} "3";"5"
   \ar@{-}|{} "4";"5"
   \ar@{-}|{} "1";"4"
   \ar@{-}|{} "4";"2"
   \ar@{-}|{} "2";"3"
   \ar@{-}|{} "3";"1"
 \end{xy}
 \end{equation*}
 and charges 
 \begin{equation}
  Q = \left(\begin{array}{ccccc}
                  1 & 1 &  0 &  0 & -N \\
                  1 & 1 & -1 & -1 & 0 
               \end{array}
         \right).
 \end{equation}
 For $N >2$, the model has a unique IR phase (the smooth phase) and two UV phases.  
 Performing the instanton sums in one of the UV phases, we find
 \begin{equation}
  F(\omega) = -\frac{1}{2 N \omega\left(1+\omega\right)}.
 \end{equation}
\end{itemize}
The reader can easily verify that in these three seemingly different cases, 
the $F(\omega)$ computed from the instanton expansion is precisely the 
$F(\omega)$ computed by eqn. (\ref{eq:F}).  

These results lead us to a conjecture:  Given a two-parameter GLSM  with a 
standard basis of charges, $\Delta \neq 0$, and an irreducible  
(over $\Z$) $P(\omega)$, the non-zero $A$-model correlators are given by 
\begin{equation}
Y_{a,b} = \sum_{\hat{\omega} | P(\hat{\omega})} 
          s(\hat{\omega})^{n_1} \hat{\omega}^b  F(\hat{\omega}), 
          ~~~\text{for}~~a+b = d + \Delta n_1,
\end{equation}
where 
\begin{eqnarray}
s(\omega) & = & \mu^\Delta q_1(\mu) \prod_i \zeta_i^{-Q^1_i} \nonumber\\
F(\omega)  & = & \frac{\sign(\Delta)}{\sum_i \left(Q_i^2\right)^2 \prod_{j\neq i} \zeta_j},
\end{eqnarray}
and $\zeta_i = Q^1_i + Q^2_i \omega$.

The GLSM is known to be democratic in certain aspects---for example, 
a Calabi-Yau model seems to contain very different phases:
some may have a nice geometric interpretation, others may be Landau-Ginzburg
theories, and yet others may be exotic mixtures of the two.  Despite this 
diversity before our undiscerning eyes, the GLSM shows us that certain aspects, 
like the chiral ring, remain independent of the particular phase.  The 
seemingly universal form of $F(\omega)$ is another manifestation of the GLSM's 
democratic principles.

\section{Discussion} \label{s:Discussion}
We have spent the bulk of the paper exploring the properties of $A$-model
observables in two-parameter GLSMs.  We have found that these observables
are supported by the Higgs branch in the UV, and we have shown that, while
they cannot be supported by the Higgs branch in IR, it is plausible that
they are supported by the isolated Coulomb vacua.   In this section, we
would like to discuss this and related findings from several perspectives,
including a possible space-time interpretation for our two-dimensional
findings.

\subsection{Instanton sums in ``wrong'' phases}
We have seen that in models with $\Delta \neq 0$  the ghost number selection
rule (eqn. (\ref{eq:ghnum})) ensures that, while the UV phase Higgs branch
supports an infinite number of correlators, the IR Higgs branch cannot support
most of these.  Of course, as the astute reader has noted, the $n_1=0$ correlators
are not excluded by this rule, and we could attempt to calculate them 
in other phases.  Let us consider the case of the smooth phase in the
example of $\Orbt$ studied in section \ref{s:Example}. It is easy to see 
that here all instantons with $n_2 \ge 0$ should contribute.
The standard techniques of Morrison and Plesser determine the 
$n_2 >0$ contributions, while the $n_2 =0$ contributions are fixed by assuming 
that the correlators satisfy the quantum cohomology relations of 
eqn. (\ref{eq:mrexample}).  The final result is that the computations of 
the $n_1=0$ instanton sums in the
smooth phase differ by a sign from the $n_1=0$ instanton sums in the IR
phase.  In a sense, a difference of this sort should not surprise us:
we know that there are Coulomb vacua in this phase, and the instanton
sum cannot be reliable.  In fact, it is more surprising that the difference is
one of merely a minus sign.  In many other cases, like that of the 
$\C^3/\Z_{3(211)}$ model mentioned above, the difference is more drastic.  
There, an instanton computation of the $Y_{3-b,b}$ correlators in the IR
phase would only have contributions from $\nv =0$ instantons, yet
the UV computation results in $Y_{3-b,b}$ with a non-trivial $q_2$ dependence.

This suggests that all of the topological correlators, including those
that are ``allowed'' to be calculated on the IR Higgs branch are, in fact,
supported on the IR Coulomb branch. 

In addition, one might be tempted to try to compute instanton sums in
intermediate phases. We have not studied this in detail, but some 
preliminary investigations have suggested that here the instanton 
sums lead to expressions that are non-rational functions of the $q_a$.
It seems that this can again be attributed to the existence of Coulomb
vacua in these phases.

\subsection{Singularities of Coulomb Branch Correlators}
Although we have shown that an instanton computation is inappropriate in
IR phases, and we have attributed this to the emergence of the Coulomb
branch, the reader has surely noted that we have not performed a 
calculation on the Coulomb branch to verify our findings.  Instead,
we argued that the form of the correlators that follows from the
quantum cohomology relations is consistent with the expected form
of a Coulomb branch computation. 

The Higgs computation of the correlators had a very nice feature:
the singularities in the correlators were associated to a reasonable
physical phenomenon:  the un-Higgsing of a gauge group and the emergence
of the continuous Coulomb branch.  Can we make a similar association
for the isolated Coulomb computation of the correlators?  This is
particularly urgent if, as in the previous section, we argue that
even the $n_1 =0$ correlators cannot be computed by an instanton
sum  in the IR.

Restricting attention to the $u=2$ case, it is clear that
singularities in the Higgs correlators are due to the un-Higgsing
of the gauge subgroup with charges $Q^2_i$, while keeping $\sigma_1$
massive.  This will occur when at least one matter multiplet is
uncharged under this subgroup.  In that case, the semi-classical
prediction of this singularity places it at
\begin{equation}
q_2^\ast =  \prod_{i|Q^2_i\neq 0} \left(Q^2_i\right)^{Q^2_i}.
\end{equation}
As $q_2$ approaches $q_2^\ast$,  then some of the isolated Coulomb 
vacua have the property that $\omega$ is sent to $\infty$, and, since 
$s(\omega) \sim \omega^{-\Delta}$ for large $\omega$, we see that
$\sigma_1 \sim \omega^{-1}$ and $\sigma_2 \sim \text{const}$.

Recall, that the continuous Coulomb branch has  $\sigma_1 = 0$,
and it can be reliably studied by the effective twisted superpotential
when $\sigma_2$ is large.   If the continuous Coulomb branch continues 
to finite $\sigma_2$, then is seems sensible to attribute  the 
singularity in the correlators to some of the isolated vacua approaching
the continuous Coulomb branch.

The isolated Coulomb vacua can obviously have a different type of
seemingly pathological behavior:  as $q_2$ is varied, some of the
isolated vacua will merge.  This will happen when 
$\discrim_\omega P(\omega) = 0.$  
Since $F(\omega) \sim P'(\omega)^{-1}$, it is clear that near
these $q_2$ the weighting factor $Z(\sigma_1,\sigma_2)$ will
diverge, yet these divergences do not show up
in the $A$-model correlators.\footnote{To be precise, this is
true for $q_2 \neq 0, \infty$.}It would be interesting
to understand why the merging of isolated and continuous
Coulomb branches seems to lead to singularities in correlators, 
while the merging of two isolated Coulomb vacua does not lead
to singularities in the correlators.

\subsection{General $A$-model correlators and {\protect $F(\omega)$} }
We have made some progress towards computing $A$-model 
correlators in general toric GLSMs.   Our analysis of 
GLSMs with a general orbifold phase covers all two-parameter
models with applications to localized tachyon condensation.
Furthermore, if our conjecture for the form of $F(\omega)$ 
is correct, then we have determined the $A$-model correlators
in all two-parameter models.  

It would also be very useful to generalize the analysis beyond
$u=2$.  Two parameter models are quite special:  for example, once
the standard basis is chosen for $Q^a_i$, $Q^2_i$ are essentially
unique.  Even more fundamentally, all cones in $\R^2$ are simplicial,
which greatly simplifies the structure of the instanton sums.
While much of our approach is tractable precisely because $u=2$,
it seems possible that more general techniques like Gr\"obner bases
and elimination theory may still be tractable due to the underlying
toric structure of the problem\cite{Sturmfels:polys}.

\subsection{Towards a Space-time Interpretation}
Although properties of RG flows in $d=2$ theories are certainly
interesting, the real interest of this project lies in the 
space-time interpretation of the world-sheet physics.  Unfortunately,
this is difficult for several reasons.  First, although the chiral
ring of the GLSM is readily identified with the chiral ring of
the UV orbifold theory, and the quantum cohomology relations 
of the GLSM correspond to the deformed chiral ring relations of
the orbifold, due to the absence of space-time supersymmetry, 
it remains difficult to understand what precisely the chiral
ring is computing in space-time.  The second, and perhaps more
serious, problem stems from the fact that, unlike in the case
of open string tachyons, the IR fixed point seen in the GLSM 
RG flow is certainly {\em not} the endpoint of tachyon 
condensation \cite{MNP:loctac}.  At best, it should be thought of as a description 
suitable for many decades of the RG time.  Thus, we cannot
match the chiral ring that we track through the flow to some
sensible static space-time physics in the IR.  Instead, we
must try to interpret our results in some non-static background.
This seems perilous indeed, but let us attempt it.

The interpretation that the GLSM suggests is that the IR 
phase corresponds to an expanding space-time, where divisors are getting 
large, and any non-trivial $\alpha'$ effects are suppressed.  If we 
trust this picture, it seems natural to suggest that the Coulomb vacua 
and the chiral ring they seem to support should be associated to 
the ``edge'' of this expanding space-time.  This would easily explain
why these effects are difficult to interpret from the point of view
of the IR Higgs branch, which, one would think, should be associated
to the physics discernible to an observer inside the expanding bubble.
This interpretation is consistent with the decoupling of the two IR 
branches that could be seen from a naive classical argument that 
neglected K\"ahler term renormalization.

If this interpretation is reasonable, then it is clear that the
difficulties of interpreting the Coulomb branch physics are tied
to the non-compactness of the models we have been considering.
It would be interesting to construct models where our framework
describes some local geometry in a compact space.  One could
hope that with a suitable set of limits, one would be able to
describe the physics of the Coulomb branch  in terms of the
compact space.

While such efforts are worthwhile, and may at least reveal whether
the proposed connection is on the right track, they
will still be within the limitations of the RG approach to tachyon
condensation:  while providing a heuristically reasonable picture,
there are great difficulties in relating the RG flow to space-time
physics.  As we have shown, the RG flow of the GLSM contains a
lot of rich physics, and topological methods can be used to probe
these in a quantitative fashion.  New methods are needed to go
beyond heuristics in translating these world-sheet riches into
space-time gains.
%%%%%%%%%%%%%%%%%%%%%%%%%%%%%%%%%%%%%%%%%%%%%%%%%%%%%%%%%%%%%%%%
\section*{Acknowledgments}
It is a pleasure to thank R.~Duivenvoorden, C.~Haase, E.~Martinec, G.~Moore, D.~Morrison 
and S.~Rinke for useful comments and conversations.  This article is based upon 
work supported in part by the National Science Foundation under Grants 
DMS-0074072 and DMS-0301476.  Any opinions, findings, and conclusions or 
recommendations expressed in this article are those of the authors and 
do not necessarily reflect the views of the National Science Foundation.

%%%%%%%%%%%%%%%%%%%%%%%%%%%%%%%%%%%%%%%%%%%%%%%%%%%%%%%%%%%%%%%%
\appendix
\section{Some Toric Geometry} \label{app:toric}
To make our work more self-contained, we provide some additional details about 
the structure of the instanton moduli spaces $\Mn$.

Let $d_i = \la \nv, \Qv_i \ra$ and $I_{\nv} = \left\{i | d_i <0\right\}$.
Elaborating on eqn. (\ref{eq:Mnform}),  $\Mn$ is a toric variety of dimension
$\dim(\Mn) =  \sum_{i\not\in I_{\nv}} d_i - u$, 
defined by the homogeneous coordinates $\xi_{ij}$, 
$ i \not\in I_{\nv}$, $j=0,\ldots,d_i$ with charges $Q_i^a$ under 
$\left(\C^*\right)^u$ and an excluded set $F_{\nv}$.  The intersection
ring of $\Mn$ is generated by classes $\sigma_a^{\nv}$, in terms of which
the divisor $\xi_{ij} = 0$ is
\begin{equation}
\xi_i^{(\nv)} = \sum_a Q_i^a \sigma_a^{(\nv)}.
\end{equation}
We will use the notation $\xi_i^{(\nv)}$ for this linear combination of
generators of the intersection ring even for $i\in I_{\nv}$.  Note that (up to
scaling by $\mu$) these
divisors are obtained when, following section \ref{s:toptwist}, 
 $\xi_i$ is mapped as a polynomial in $\sigma_a$ to a divisor in $\Mn$.

$F_{\nv}$ may be complicated, but a certain subset of it is determined
by the cone $\K$.  $\K$ may be presented as the set of points in $\R^u$ satisfying
a set of inequalities: 
\begin{equation}
\K = \left\{\rv \in \R^u | \la \mv^\alpha, \rv\ra > 0, \alpha =1 \ldots A\right\}.
\end{equation}
Then 
\begin{equation}
 \left\{\xi_{i~1} = \xi_{i~(d_i+1)}=0 | 
   \la \mv^\alpha, Q_i \ra > 0 ~~\text{and}~~ i\in I_{\nv}\right\} \subset F_{\nv}.
\end{equation}
As explained in \cite{MP:summing}, each of these $A$ intersections of hyperplanes leads
to a (Stanley-Reisner) relation on the intersection ring of $\Mn$:
\begin{equation}
\label{eq:SRrel}
 \prod_{ \genfrac{}{}{0pt}{}{d_i\ge 0}{\la\mv^\alpha, Q_i\ra > 0}} 
  \left(\xi_i^{(\nv)}\right)^{d_i + 1} = 0.
\end{equation}
When little or no confusion can arise,
we will drop the $\nv$ superscripts on the divisors $\xi_i^{\nv}$.

Modulo the nontrivial scaling by $\mu$ discussed in section \ref{s:toptwist},
the class ${\cal O}^{(\bf n)}$ is obtained from $\cal O$ expressed as a
polynomial in $\sigma_a$ by writing the same polynomial in
$\sigma_a^{({\bf n})}$.  

The Euler class is constructed from the divisors $\xi_i^{\nv}$ with $d_i <0$:
\begin{equation}
\ECn = \prod_{i\in I_{\nv}} \left(\xi_i^{({\bf n})}\right)^{-1-d_i}.
\end{equation} 
Its degree is $\sum_{i\in I_{\nv}} (-1-d_i)$.  One can check that this is 
consistent with the ghost number selection rule:  unless 
\begin{equation}
\deg(\O) + \sum_{i\in I_{\nv}} (-1 -d_i) = \dim(\Mn) = \sum_{i \not\in I} (d_i + 1)  - u,
\end{equation}
instantons with instanton number $\nv$ do not contribute to $Y_{\O}$.

\section{An instanton computation} \label{app:instsum}
The reader may be interested in how we obtain eqns. (\ref{eq:yexmpl}, \ref{eq:ygen}).
In this appendix we will fill in some of the steps of this computation.  We will
work with the example of $\Orbt$.  First, we will show how we may derive a formula
for the instanton contribution $Y_{a,b}^{\nv}$, and then we will show how the 
instanton sum can be manipulated into the form of eqn. (\ref{eq:ygen}).  We will
use the results and notation of appendix \ref{app:toric}.

To sum the instantons in the orbifold phase of $\Orbt$, we begin by determining the
dual cone $\Kd$ of contributing instantons $n$.  $\K$ determines $\Kd$ to be the
cone shown in figure $\ref{fig:kdexmpl}$.
\begin{figure}
\[
\begin{xy} <1.0mm,0mm>:
  (0,0)*{\bullet} ="0", (12,0)*{}="1", (0,40)*{}="2", (-50,25)*{}="3", 
  (-20,-40)*{}="4", (-50,0)*{}="5", (0,-50)*{}="6", 
  (10,-4)*{n_1}, (-4,38)*{n_2}, (-30,-10)*{\Kd},
  (-25,12.5)*{\bullet},  (-19,15)*{\left(\genfrac{}{}{0pt}{}{-2}{1}\right)}, 
  (-15, -30)*{\bullet}, (-10,-30)*{\left(\genfrac{}{}{0pt}{}{-1}{-N}\right)},
  \ar@{-}|{} "0"; "3"
  \ar@{-}|{} "0"; "4"
  \ar@{.}|{} "1"; "5" 
  \ar@{.}|{} "2"; "6" 
\end{xy}
\]
\caption{The dual cone in the orbifold phase of {\protect $\Orbt$}.}
\label{fig:kdexmpl}
\end{figure}
Thus, the contributing instantons have $n_1 < 0$ and 
$ N n_1 \le n_2 \le \lfloor -\frac{n_1}{2} \rfloor$.  Since
\begin{equation}
d_i = \left( n_1, n_1, n_1+n_2, n_2 - N n_1, -n_1-2 n_2\right),
\end{equation}
we see that 
\begin{equation}
F_{\nv} = \left\{\xi_{4j} = 0| j=0,\ldots,d_4\right\} \cup
          \left\{\xi_{5j} = 0| j=0,\ldots,d_5\right\}.
\end{equation}
This leads to the Stanley-Reisner relations on the intersection ring
of $\Mn$:
\begin{eqnarray}
\left(\sigma_2^{(\nv)} - N \sigma_1^{(\nv)} \right)^{d_4+1} & = & 0,\nonumber\\
\left(-\sigma_1^{(\nv)} - 2\sigma_2^{(\nv)} \right)^{d_5+1}    & = & 0.
\end{eqnarray}
We may also write down the Euler class:
\begin{equation}
\ECn = \prod_{i\le3} \left(\xi_i^{(\nv)}\right)^{-d_i -1}.
\end{equation}
We will henceforth drop the superscript $(\nv)$ on the various divisors.
The $\sigma_1,\sigma_2$ basis for the divisors is a bit inconvenient for the
purposes of writing down the $Y_{a,b}^{\nv}$.  A more convenient basis is
\begin{equation}
\left( \begin{array}{c} \lambda_1 \\ \lambda_2 \end{array} \right) =
\left( \begin{array}{c} \xi_4 \\ \xi_5 \end{array}\right) =
\left( \begin{array}{cc} -N & 1 \\ -1 & -2 \end{array} \right)
\left( \begin{array}{c} \sigma_1 \\ \sigma_2 \end{array}\right).
\end{equation}
In this basis the Stanley-Reisner relations are $\lambda_1^{d_4 +1} = 0$
and $\lambda_2^{d_5 +1} = 0$, and the moduli space is 
$\Mn = \P^{d_4} \times \P^{d_5}$, with the $\lambda_1, \lambda_2$
being the fundamental classes on the $\P$s.  Writing the Euler class and the
$\sigma_1,\sigma_2$ in terms of the $\lambda_1,\lambda_2$, eqn. (\ref{eq:yn}) becomes
\begin{eqnarray}
Y_{a,b}^{\nv} & = & \frac{1}{M} \la \sigma_1^a \sigma_2^b \ECn\ra_{\Mn} \nonumber\\
 ~ & = & \frac{1}{M} \la \left(\frac{2\lambda_1 + \lambda_2}{-M}\right)^{a-2 n_1 -2}
                         \left(\frac{-\lambda_1 + N \lambda_2}{-M}\right)^b
                 \left(\frac{\lambda_1 + (N+1)\lambda_2}{-M}\right)^{-n_1-n_2-1} \ra_{\Mn}.
\end{eqnarray}
We have let $M = 2N+1$, and the leading factor of $\ff{1}{M}$ is due to over-counting
the instantons because of the remaining discrete gauge symmetry in the orbifold phase.
The intersection theory on $\Mn$ is extremely simple, and $Y_{a,b}^{\nv}$ is just
the coefficient of $\lambda_1^{d_4}$ in the above product of binomials.  Note
that if $a+b = 3+ (2-N) n_1$, then the power of $\lambda_2$ is correct, and
otherwise $Y_{a,b}^{\nv}$ vanishes.  By letting $\lambda_1 = 1$ and  $\lambda_2 = z$,
we can write the coefficient of $\lambda_1^{d_4}\lambda_2^{d_5}$ as
a contour integral about $z=0$.  Trivial manipulations then yield eqn. (\ref{eq:yexmpl}),
which we reproduce here for convenience:
\begin{eqnarray}
\label{eq:yexmpl2}
Y_{a,b}^{n_1,n_2}& =& \mu^{n_1 (2-N)} \ff{1}{M} \oint\limits_{C(0)} \frac{dz}{2\pi i} 
                    \frac{ (2+z)} {z (1 +(N+1)z)}
                    \left[ \frac{ (-M)^{N+1} z }{ (2+z)^N (1+(N+1)z)}\right]^{n_1}
                    \times\nonumber\\
                 &~& ~~~~~~~~~ \times\left[ \frac{-M z^2}{1+(N+1)z} \right]^{n_2}
                    \left[ \frac{-1+Nz}{2+z} \right]^b.
\end{eqnarray}

We will now work further with this explicit form to show in an example 
how to obtain eqn. (\ref{eq:ygen}).   We begin by performing the sum on $n_2$, 
which runs for $ N n_1 \le n_2 \le \lfloor - \frac{n_1}{2} \rfloor$: 
\begin{eqnarray}
Y_{3+(2-N)n_1-b,b} & =& \left(\mu^{(2-N)} q_1\right)^{n_1} \ff{1}{M} \oint\limits_{C(0)} \frac{dz}{2\pi i} 
                    \frac{ (2+z)} {z (1 +(N+1)z)}
                    \left[ \frac{ (-M)^{N+1} z }{ (2+z)^N (1+(N+1)z)}\right]^{n_1}
                    \times\nonumber\\
                 &~& ~~~~~~~~~ \times \left[ \frac{-1+Nz}{2+z} \right]^b
                           \frac{ R^{N n_1} - R^{\lfloor -\frac{n_1}{2} \rfloor+1} }{1- R}.
\end{eqnarray}
where $R = -M z^2 q_2 /( 1+(N+1)z )$.  The term with $R^{\lfloor -\frac{n_1}{2}\rfloor +1}$ 
does not contribute to the residue at zero, so we may discard it.  Now, we change coordinates
to
\begin{equation}
w = \frac{-1+Nz}{2+z}.
\end{equation}
This leads to a form familiar from eqn. (\ref{eq:ygen}):
\begin{equation}
Y_{3+(2-N)n_1-b, b} = \left(\mu^{2-N}q_1\right)^{n_1} \oint\limits_{C\left(-\frac{1}{2}\right)} \frac{dw}{2\pi i}
                     (-1-2w)^{(2N+1)n_1 -1} (1+w)^{-(N+1)n_1} q_2^{N n_1} \frac{w^b}{ P(\omega)}.
\end{equation} 
It is evident that the poles of the integrand are $\omega=-1/2$ and the roots of $P(\omega)$, 
so, finally, we find
\begin{equation}
Y_{3+(2-N)n_1-b,b} = \mu^{N-2}q_1(\mu)^{-1} \sum_{\hat{\omega}|P(\hat{\omega})}
               ~\oint\limits_{C\left(\hat{\omega}\right)} \frac{dw}{2\pi i} 
        \left( \frac{(-1-2\omega) (w-N)^N}{ 1+\omega}\right)^{n_1} \frac{w^b}{(1+2\omega) P(\omega)}.
\end{equation}
Comparing to the standard form of the correlators, $F(\omega) = \left((1+2\omega) P'(\omega)\right)^{-1}$,
which at the roots of $P(\omega)$ reduces to the familiar 
$F(\omega) = \left(1+3N+2N\omega\right)^{-1}$.

\section{A Proof of Quantum Cohomology Relations} \label{app:mrproof}
In this appendix we prove that eqn. (\ref{eq:mrexpl}) holds, provided that
the $A$-model correlators may be computed by the standard instanton 
sums in {\em some} phase $\K$.  We will use notation given in the text as
well as appendix \ref{app:toric}.

Expanding eqn. (\ref{eq:mrexpl}) in powers of $q_a$, and using eqn. (\ref{eq:yn}), we
find that eqn. (\ref{eq:mrexpl}) is equivalent to three statements.
The first of these,
\begin{equation}
\label{eq:cond1}
\la \O \PPROD \EUL \ra_{\Mn}  = \la \O \MPROD \MEUL \ra_{\M_{\nv - \ev_a}},
\end{equation}
must hold when $\nv$ and $\nv-\ev_a$ are both in $\Kd$.  The second and 
third should hold  when one of the vectors is in $\Kd$ while the other is not:
\begin{eqnarray}
\la \O \PPROD \EUL \ra_{\Mn} & = & 0, ~~~\nv \in \Kd, \nv-\ev_a \not\in \Kd,  \\
\la \O \MPROD \MEUL \ra_{\M_{\nv-\ev_a}} & = & 0, ~~~\nv \in \Kd, \nv-\ev_a \not\in \Kd.
\end{eqnarray}

To prove the first of these conditions, we will now establish some additional 
properties of the intersection ring on $\Mn$.
Suppose $\nv,\nv-\ev_a \in \Kd$.  Furthermore, consider the case that $I_{\nv} = I_{\nv-\ev_a}$.
Then for any $\U$, 
\begin{equation}
\la \U \prod_{\genfrac{}{}{0pt}{}{Q_i^a > 0}{d_i \ge 0}} \xi_i^{Q_i^a} \ra_{\Mn} = 
\la \U \prod_{\genfrac{}{}{0pt}{}{Q_i^a < 0}{d_i \ge 0}} \xi_i^{-Q_i^a}\ra_{\M_{\nv - \ev_a} }.
\end{equation}
This follows because $d_i(\nv-\ev_a) = d_i - Q_i^a$ and the intersection of
$Q_i^a$ copies of $\xi_i$ in $\Mn$ effectively leaves a variety described
by $d_i -Q_i^a$ copies of $\xi_i$---$\M_{\nv-\ev_a}$.

More generally, $I_{\nv} \neq I_{\nv- \ev_a}$, and the above is generalized to
\begin{equation}
\label{eq:basicrel}
\la \U  \prod_{\genfrac{}{}{0pt}{}{Q_i^a > 0}{d_i       \ge 0}} \xi_i^{\min(Q_i^a, d_i+1)} \ra_{\Mn} = 
\la \U  \prod_{\genfrac{}{}{0pt}{}{Q_i^a < 0}{d_i-Q_i^a \ge 0}} \xi_i^{\min(-Q_i^a, d_i+1-Q_i^a)}
\ra_{\M_{\nv-\ev_a}}.
\end{equation}
To see this, observe that if $Q_i^a > d_i \ge 0$, then $\M_{\nv-\ev_a}$ contains
{\em no} copies of $\xi_i$, which is achieved by restricting to the intersection of
$d_i+1$ copies of $\xi_i$ in $\Mn$.

A few moments' thought will convince the reader that
\begin{equation}
\label{eq:ugly}
\prod_{Q_i^a>d_i\ge 0}\xi_i^{Q_i^a-d_i-1}
\prod_{\genfrac{}{}{0pt}{}{Q_i>0}{d_i<0}}\xi_i^{Q_i^a} \prod_{d_i<0}\xi_i^{-d_i-1} = 
\prod_{d-Q_i^a<0}\xi_i^{Q_i^a-d_i-1} \prod_{\genfrac{}{}{0pt}{}{Q_i^a<0}{d_i-Q_i^a<0}}
\xi_i^{-Q_i^a} \prod_{\genfrac{}{}{0pt}{}{d_i-Q_i^a\ge 0}{ d_i<0}} \xi_i^{-d_i-1}\ .
\end{equation}

We are finally in a position to prove our claim.  Consider 
\begin{equation}
\U = \O \prod_{Q_i^a > d_i \ge 0} \xi_i^{Q_i^a - d_i -1}
        \prod_{\genfrac{}{}{0pt}{}{Q_i^a > 0}{d_i <0}} \xi_i^{Q_i^a} 
        \EUL.
\end{equation}
On one hand, 
\begin{equation}
\U \prod_{\genfrac{}{}{0pt}{}{Q_i^a > 0}{d_i \ge 0}} \xi_i^{\min(Q_i^a,d_i+1)} = \O \PPROD \EUL.
\end{equation}
On the other hand, using eqn. (\ref{eq:ugly}) in $\U$, we have
\begin{equation}
\U  \prod_{\genfrac{}{}{0pt}{}{Q_i^a <0 }{d_i - Q_i^a \ge 0}} \xi_i^{\min(-Q_i^a,d_i+1-Q_i^a)} 
=  \O \MPROD \MEUL.
\end{equation}
Using  eqn. (\ref{eq:basicrel}), we see that eqn. (\ref{eq:cond1}) holds.

The second and third conditions are easier. Construct $V$, the toric variety
associated to $\K$ as a K\"ahler quotient.  The form of the $D$-terms (the moment map
in geometric terminology) shows that the cone $\K$ is generated by a subset of the 
$\Qv_i$, in fact, precisely those $\Qv_i$ that satisfy 
$\la\mv^\alpha, \Qv_i\ra >0$ for $\alpha = 1,\ldots,A$.\footnote{Recall that $\mv^\alpha$
determined a set of inequalities that defined $\K$.}   Thus, if $\nv \in \Kd$, then for any
$i$ such that $\la \mv^\alpha, \Qv_i\ra >0$ for all $\alpha$, $d_i = \la n,\Qv_i\ra \ge 0$.
If, in addition, $\nv -\ev_a \not\in \Kd$, then there exists some $i$ such that
$\la \mv^\alpha, \Qv_i \ra >0$ and  $d_i - Q_i^a = \la \nv - \ev_a, Q_i\ra < 0$.  So, we
conclude that if $\nv \in \Kd$ and $\nv-\ev_a \not\in \Kd$, then there exists some
$i$ such that $\la \mv^\alpha, \Qv_i \ra >0$, and $Q_i^a > d_i \ge 0$.   The second
condition then follows as a consequence of the corresponding Stanley-Reisner
relation in eqn. (\ref{eq:SRrel}).  The third condition is shown to hold in the same
manner.

%\bibliographystyle{unsrt}
%\bibliography{big}

\end{document}